\documentclass[aps,twocolumn,prd,showpacs,nofootinbib]{revtex4}

\usepackage{amsmath}
\usepackage{graphicx}
\usepackage{amssymb}

\newcommand{\gta}{\gtrsim}

\newcommand{\dd}{\mathrm{d}}

\newcommand{\cs}{c_{_{\mathrm{S}}}}

\newcommand{\nS}{n_{{\mathrm{s}}}}
\newcommand{\alphaS}{\alpha_{{\mathrm{s}}}}

\newcommand{\uini}{\mathrm{ini}}
\newcommand{\ue}{\mathrm{e}}

\newcommand{\uk}{\mathrm{k}}
\newcommand{\us}{\mathrm{s}}

\newcommand{\Tcmb}{T_\mathrm{cmb}}

\newcommand{\calH}{\mathcal{H}}
\newcommand{\calP}{\mathcal{P}}

\newcommand{\mpl}{m_{_\mathrm{Pl}}}

\newcommand{\ie}{\textsl{i.e.}~}

\newcommand{\bear}{\begin{array}}
\newcommand{\eear}{\end{array}}
\newcommand{\bea}{\begin{eqnarray}}
\newcommand{\eea}{\end{eqnarray}}
\newcommand{\beq}{\begin{equation}}
\newcommand{\eeq}{\end{equation}}
\newcommand{\bef}{\begin{figure}}
\newcommand{\eef}{\end{figure}}
\newcommand{\bec}{\begin{center}}
\newcommand{\eec}{\end{center}}

\newcommand{\abs}[1]{\vert{#1}\vert}

\newcommand{\order}[1]{\mathcal{O}\!\left(#1\right)}

\newcommand{\deriv}[2]{#1_{\negthinspace,#2}}

\newcommand{\Airy}[1]{\mathrm{Ai}\!\left(#1\right)}
\newcommand{\Biry}[1]{\mathrm{Bi}\!\left(#1\right)}
\newcommand{\besselJ}[1]{J{_{#1}}}

\newcommand{\ellF}[2]{F\!\left(#1 \left|#2\right.\right)}
\newcommand{\ellE}[2]{E\!\left(#1 \left|#2\right.\right)}

\newcommand{\ud}{\mathrm{d}}

\newcommand{\calHS}{\mathcal{H_*}}
\newcommand{\etaS}{\eta_*}

\newcommand{\piv}{\diamond}
\newcommand{\pivhub}{\triangleright}
\newcommand{\etaP}{\eta_\piv}

\newcommand{\epsoneP}{\epsilon_{1\piv}}
\newcommand{\epstwoP}{\epsilon_{2\piv}}
\newcommand{\epsthreeP}{\epsilon_{3\piv}}
\newcommand{\epsfourP}{\epsilon_{4\piv}}
\newcommand{\deltaoneP}{\delta_{1\piv}}
\newcommand{\deltatwoP}{\delta_{2\piv}}
\newcommand{\deltathreeP}{\delta_{3\piv}}

\newcommand{\csP}{\cs {}_\piv}
\newcommand{\epsoneS}{\epsilon_{1*}}
\newcommand{\epstwoS}{\epsilon_{2*}}

\newcommand{\deltaoneS}{\delta_{1*}}

\newcommand{\epsonePH}{\epsilon_{1\pivhub}}
\newcommand{\etaPH}{\eta_\pivhub}

\newcommand{\gs}{g_\us}

\newcommand{\alphas}{\alpha'}

\newcommand{\IR}{_{_{\mathrm{IR}}}}
\newcommand{\UV}{_{_{\mathrm{UV}}}}
\newcommand{\fNL}{f_{_{\mathrm{NL}}}}

\begin{document}

\title{K-inflationary Power Spectra in the Uniform Approximation}

\author{Larissa Lorenz} \email{lorenz@iap.fr}
\author{J\'er\^ome Martin} \email{jmartin@iap.fr} \affiliation{
Institut d'Astrophysique de Paris, UMR 7095-CNRS, Universit\'e Pierre
et Marie Curie, 98bis boulevard Arago, 75014 Paris, France}

\author{Christophe Ringeval}
\email{christophe.ringeval@uclouvain.be}
\affiliation{Theoretical and Mathematical Physics Group, Centre for
  Particle Physics and Phenomenology, Louvain University, 2 Chemin du
  Cyclotron, 1348 Louvain-la-Neuve, Belgium}

\date{January 7, 2009}

\begin{abstract}
  The advent of explicit Dirac--Born--Infeld (DBI) inflationary models
  within string theory has drawn renewed interest to the cosmological
  role of unusual scalar field dynamics, usually referred to as
  k-inflation. In this situation, the standard method used to
  determine the behavior of cosmological perturbations breaks down.
  We present a generic method, based on the uniform approximation, to
  analytically derive the power spectra of scalar and tensor
  perturbations. For this purpose, a simple hierarchy of parameters,
  related to the sound speed of the cosmological fluctuations and its
  successive derivatives, is introduced in a k-inflation analogue of
  the Hubble flow functions. The scalar spectral index and its running
  are obtained up to next to next to leading order for all
  k-inflationary models. This result relies on the existence
  of a well-motivated initial state, which is not trivial in the
  present context: having the wavelength of the Fourier mode smaller
  than the sonic horizon is indeed not enough and some conditions on
  the dynamics of the sound speed are also required. Our method is
  then applied to various models encountered in the literature.  After
  deriving a generic slow-roll trajectory valid for any DBI model,
  simple formulae for the cosmological observables are obtained. In
  particular, the running, as the spectral index, for the so-called UV
  and IR brane inflationary models is found to be uniquely determined
  by the 't~Hooft coupling. Finally, the accuracy of these
  cosmological predictions is assessed by comparing the analytical
  approximations with exact numerical integrations.
\end{abstract}

\maketitle

\section{Introduction}
\label{sec:intro}

The primordial matter perturbations held responsible for growth and
formation of large scale structure are commonly traced back to quantum
fluctuations of a scalar field $\varphi$ which should have dominated
the energy density in the Universe at early times. The primordial
power spectrum of both the scalar and tensor perturbations is a 
calculational output of the inflationary scenario, and the recent
Cosmic Microwave Background (CMB) experiments have gathered
considerable evidence in favor of it~\cite{Gold:2008kp, Hill:2008hx,
  Hinshaw:2008kr, Nolta:2008ih, Dunkley:2008ie, Komatsu:2008hk}.

\par

A compelling virtue of the inflationary paradigm is the fact that it
can be sustained by a whole class of scalar field potentials
$V(\varphi)$, provided these exhibit characteristics (\ie in their
slope and curvature) in agreement with the ``slow-roll'' conditions:
while inflation is under way, the potential must dominate over the
kinetic energy $\dot{\varphi}^{2}/2$, that is, $\ln V$ should be flat
enough not to accelerate the field quickly. The slow-roll regime for
inflation is described by a hierarchy of parameters $\epsilon_{i}$,
assumed to be small, in which the primordial power spectra can be
analytically expressed as a Taylor expansion~\cite{Stewart:1993bc,
  Martin:1999wa, Martin:2000ak, Schwarz:2001vv, Leach:2002ar,
  Schwarz:2004tz}.

\par

A large body of  literature is devoted to the search for inflationary
scenarios, in the sense that they should be naturally motivated by a
high energy physics theory~\cite{Lyth:1998xn}. For example, the toolkit
of ten-dimensional super-string theory and its various compactifications
to an effective four-dimensional field theory has been used to design
candidate potentials~\cite{Cline:2006hu,
Kallosh:2007ig,McAllister:2007bg}.

\par

An interesting approach studies the effect of modifications to the
kinetic term of the inflaton
(k-inflation)~\cite{ArmendarizPicon:1999rj}; in the perturbation
treatment, these modifications manifest themselves as a (possibly
time-dependent) ``speed of sound'' $\cs\neq1$ (the speed of light being
$c=1$) for the scalar Fourier modes~\cite{Garriga:1999vw}. It turns out
that string inflation models where the inflaton field is an open string
mode are typically of this kind, with the speed of sound being a
function of the background geometry~\cite{Kachru:2003sx,
Alishahiha:2004eh}. Hence inflationary model building in string theory
combines both candidate potentials and non-canonical field evolution.

\par

The prime example of these scenarios are the so-called brane inflation
models, where the inflaton $\varphi$ corresponds to the position of a
$D$-brane within in a higher-dimensional manifold~\cite{Dvali:1998pa,
  Dvali:2001fw, Alexander:2001ks}. Being an open string mode,
$\varphi$ has a Dirac-Born-Infeld (DBI) action, which is essentially
the square root of the induced metric on the brane. This metric, in
turn, contains information about the chosen background
compactification of the extra-dimensions through the
position-dependent brane tension $T(\varphi)$. The inflaton potential
$V(\varphi)$ can be of various shape, and its exact calculation
remains a matter of active research~\cite{Baumann:2007ah,
  Baumann:2007np}. Typically, it receives contributions involving
$T(\varphi)$, but may also be affected by finer geometric detail such
as the presence of other branes in the extra-dimensional background
geometry. As a natural consequence of the non-canonical interplay
between potential and dynamics in the k-inflation case, one can no
longer trust the intuition that flat potentials support
inflation. Through $T(\varphi)$, the warping of the background acts as
a break on the field, allowing potential energy domination and
accelerated expansion of the Universe even when $\ln V$ is steep.

\par

As  it is clear from the above description, the standard methods
of slow-roll inflation break down in the case of k-inflation. In
particular, the formula expressing the background trajectory is
modified since the shape of $T(\varphi)$ (and not only the shape of
the potential as in the ordinary case) affects the motion of the
mobile brane. Another new feature of k-inflation, which is of prime
concern for this article, is that one can no longer calculate the
cosmological perturbations' power spectra using the standard
techniques. Indeed, as already mentioned above, the scalar
perturbations have now a time-dependent speed of propagation which
prevents us to integrate the equations of motion in terms of Bessel
functions. Moreover, concerning the perturbations' evolution, the
``usual'' Hubble flow functions $\epsilon_{i}$ do no longer provide a
sufficient description. Therefore, the main goal of this paper is to
put forward a general formalism for k-inflation, resembling as far as
possible the usual slow-roll formalism, where all these issues can be
addressed in a consistent and unified way.

\par

The paper is organized as follows. In Sec.~\ref{sec:dbisr}, we use a
combined hierarchy $\left(\epsilon_{i},\delta_{i}\right)$ of Hubble
and ``sound'' flow functions such that
$\epsilon_{i},\delta_{i}\ll1,i\geq1$ is the analogue of the standard
slow-roll approximation~\cite{Alishahiha:2004eh, Chen:2006nt,
  Kinney:2007ag, Peiris:2007gz}. Then, using the background equations
of motion, the slow-roll trajectory is expressed as a quadrature. Our
new formula can be applied to any DBI model (but not to
k-inflationary models in general) characterized by the functions
$T(\varphi)$ and $V(\varphi)$ and is valid under a single assumption,
namely $\epsilon _1\ll 1$. In a next step, we carry on through the
calculation of the k-inflationary perturbation spectra in full
generality (but assuming, as usual, the smallness of the Hubble and
sound flow parameters), including the non-trivial effects induced by a
varying sound speed, in the so-called uniform
approximation~\cite{Habib:2002yi, Habib:2004kc}. In particular, we
derive the scalar spectral index and the running at the next to next
to leading order for a general model of k-inflation. Moreover, we show
that a time-dependent sound horizon may lead to sub-sonic Fourier
modes starting their evolution out of the Wentzel-Kramers-Brillouin
(WKB) regime. This generically results in oscillations in the
primordial scalar power spectrum. At the end of this section, we also
discuss how the new parameters can be used to generalize to
k-inflation the classification of (single field) inflationary models
described in Ref.~\cite{Schwarz:2004tz}. Then, in
Sec.~\ref{sec:examples}, our approach is applied to various example
models encountered in the literature. In particular, we recover or
extend previous results concerning power-law DBI
inflation~\cite{Spalinski:2007dv,Spalinski:2007qy}, the
Kachru--Kallosh--Linde--Maldacena--McAllister--Trivedi (KKLMMT)
model~\cite{Kachru:2003sx}, chaotic Klebanov--Strassler (CKS)
inflation~\cite{Klebanov:2000hb, Bean:2007hc, Peiris:2007gz}, and
derive new results for models having a CKS potential plus
constant~\cite{Shandera:2006ax}. At various stages, we assess the
accuracy of our approximation scheme from exact numerical
integrations. In Sec.~\ref{sec:conclusion}, we recap and discuss our
main findings. Finally, in three short appendices, we briefly compare
our new hierarchy of parameters to the parameters already considered
before in the literature and we recall how the master equation for the
Mukhanov-Sasaki variable can be derived in k-inflation.

\section{DBI slow-roll inflation}
\label{sec:dbisr}

In this section, although we are concerned with the DBI models, the
definition of the Hubble and sound flow parameters is valid in full
generality for all k-inflationary
models~\cite{ArmendarizPicon:1999rj}. The same remark holds for the
calculation of the primordial power spectra and will be made apparent by
using explicitly the sound speed $\cs$ in the calculations.

\subsection{Basic equations}
\label{subsec:basiceq}

The basic construction behind brane inflation models in super-string
theory is discussed in several recent
reviews~\cite{Cline:2006hu,HenryTye:2006uv,Burgess:2007pz,
McAllister:2007bg}. Our starting point is the effective four-dimensional
inflaton action
\begin{equation}
\label{eq:action}
\begin{aligned}
S &= -\int {\rm d}^4x \sqrt{-g}\left[T(\varphi)
\sqrt{1+\frac{1}{T(\varphi)}\,g^{\mu \nu}\partial _{\mu}\varphi 
  \partial _{\nu}\varphi} \right. \\
& \left. \phantom{\sqrt{1+\frac{1}{T(\varphi)}}}
  +V(\varphi)-T(\varphi) \right] ,
\end{aligned}
\end{equation}
with $V(\varphi)$ the potential and $T(\varphi)$ the warp
function. The warp function is determined by a specific choice for the
geometry of the extra-dimensions. The shape of the potential receives
many different contributions, which include Coulomb-like terms
describing the attraction between branes and anti-branes as well as
terms that arise from the embedding of different dimensional
branes. On the string theory side, there is an ongoing debate on the
number and form of these contributions. The form of $T(\varphi)$ is
the subject of less controversy since known string inflation models
use the singular conifold or its cousin, the deformed
conifold~\cite{Klebanov:2000hb}. However, $T(\varphi)$ has also been
treated as a completely general function in the literature, and here,
both $V(\varphi)$ and $T(\varphi)$ will be considered as free
functions in the sake of generality.

\par

It is worth noticing that the action~(\ref{eq:action}) defines a
consistent theory. Indeed, for an arbitrary model of k-inflation, the
action of the scalar field can be written as
\begin{equation}
S_\uk=\int {\ud}^4x \sqrt{-g} P(X,\varphi),
\end{equation}
where the quantity $X \equiv -(1/2) g ^{\mu \nu}\partial _{\mu
}\varphi \partial _{\nu }\varphi$. One can show~\cite{Bruneton:2007si}
that this theory is well-defined if
\begin{equation}
  \frac{\partial P}{\partial X}>0,\qquad 
  2 X \frac{\partial ^2P}{\partial X^2}+\frac{\partial P}{\partial X}>0.
\end{equation} 
The first condition comes from the requirement that the Hamiltonian
should be bounded from below while the second is necessary if one wants
the field equations to remain hyperbolic~\cite{Bruneton:2007si}. In the
case of DBI, one has $P = -T\sqrt{1-2X/T}$ and one can check that both
conditions are indeed satisfied regardless of the sign of the brane
tension $T(\varphi)$. In the following, we will always restrict
ourselves to models of k-inflation that fulfill the above-mentioned
conditions.

\par

Variation of Eq.~(\ref{eq:action}) with respect to the metric gives the
DBI stress-energy tensor,
\begin{equation}
T_{\mu \nu}=-\frac{2}{\sqrt{-g}}\frac{\delta S}{\delta g^{\mu \nu}} \,.
\end{equation}
which is found to read
\begin{eqnarray}
\label{eq:Tmunu}
T_{\mu \nu} =\gamma
\partial _{\mu }\varphi \partial _{\nu}\varphi
-g_{\mu \nu }\left[V\left(\varphi\right)+T\left(\varphi\right)
\left(\frac{1}{\gamma}-1\right)\right],
\end{eqnarray}
where the Lorentz factor is defined
as~\cite{Silverstein:2003hf,Alishahiha:2004eh}
\begin{equation} 
\label{eq:defgamma}
\gamma \equiv \left[1+\frac{1}{T(\varphi)}
g^{\alpha \beta}\partial _{\alpha}\varphi 
\partial _{\beta }\varphi\right]^{-1/2}.
\end{equation}
The relativistic analogy becomes evident in the case where one
considers a spatially homogeneous field $\varphi(t)$ in a
Friedmann--Lema\^{\i}tre--Robertson--Walker (FLRW) universe so that
Eq.~(\ref{eq:defgamma}) simplifies to
\begin{equation}
\label{eq:gamma2}
\gamma=\frac{1}{\sqrt{1-\dot{\varphi}^{2}/T(\varphi)}}\, ,
\end{equation}
where a dot denotes a derivative with respect to cosmic time. Clearly,
$\sqrt{T(\varphi)}$ here plays the role of an upper limit on the
inflaton's velocity $\dot{\varphi}$. When expanding $\gamma$ for
$\dot{\varphi}^{2}\ll T(\varphi)$, the action at first order resumes
its canonical form. Moreover, the energy density and pressure read
\begin{equation}
\label{eq:defrhop}
\rho =\left(\gamma -1\right)T(\varphi)+V(\varphi) ,
\quad
p=\frac{\gamma -1}{\gamma}\,T(\varphi)-V(\varphi) ,
\end{equation}
so that we have the Friedmann--Lema\^{\i}tre equations
\begin{align}
\label{eq:friedmanndbi}
H^2 & = \frac{\kappa}{3}\left[(\gamma-1)T+V\right], \\
\label{eq:friedmanndbi2}
2 \dot{H} & + 3 H^2 = \kappa \left(\dfrac{1-\gamma}{\gamma} T + V \right),
\end{align}
while the Klein-Gordon equation for the field reads
\begin{eqnarray}
  \ddot{\varphi}+\frac{3H}{\gamma^{2}}\dot{\varphi}
  +\frac{3\gamma-\gamma^{3}-2}{2\gamma^{3}}\,\frac{{\rm d}T}{{\rm d}\varphi}
  +\frac{1}{\gamma^{3}}\frac{{\rm d}V}{{\rm d}\varphi}=0.
\label{eq:eofmdbi}
\end{eqnarray}
The constant $\kappa $ is defined by $\kappa\equiv8\pi/\mpl^{2}$, $\mpl$
being the four-dimensional Planck mass.

\subsection{DBI slow-roll trajectory}
\label{sec:srpara}

In standard inflation, one usually defines a hierarchy of Hubble flow
parameters from~\cite{Schwarz:2001vv, Leach:2002ar, Schwarz:2004tz}
\begin{equation}
\label{eq:defeps}
\epsilon_{n+1}=\frac{{\rm d}\ln \vert \epsilon _n \vert}{{\rm d}N}\, , 
\quad \epsilon _0\equiv \frac{H_{\mathrm{in}}}{H}\, .
\end{equation} 
Their physical interpretation is that the expansion is accelerated as
long as $\epsilon_{1}<1$ (potential energy domination). The slow-roll
approximation assumes moreover that one has
$|\epsilon_{i}|\ll1,\,i\geq1$, a condition which is in general
necessary in order to have a sufficient number of e-folds. In
addition, this last condition also allows us to integrate analytically
the field trajectory and to compute the cosmological perturbations'
power spectra.

\par

In DBI inflation, we still retain the definition~(\ref{eq:defeps}), the
only subtlety being that the Hubble parameter is now given by
Eqs.~(\ref{eq:friedmanndbi}) and~(\ref{eq:friedmanndbi2}). Expressed in
terms of derivatives of $H$ with respect to $\varphi$, the first two
Hubble flow functions read
\begin{align}
\epsilon _1 &= \frac{2}{\kappa \gamma }\frac{1}{H^2}
\left(\frac{{\rm d}H}{{\rm d}\varphi }\right)^2,\\
\epsilon _2 &= \frac{2}{\kappa \gamma }\left[\frac{2}{H^2}
\left(\frac{{\rm d}H}{{\rm d}\varphi }\right)^2 
-\frac{2}{H}
\frac{{\rm d}^2H}{{\rm d}\varphi ^2}
+\frac{1}{\gamma }\frac{{\rm d}\gamma}{{\rm d}\varphi }
\frac{1}{H}\frac{{\rm d}H}{{\rm d}\varphi }
\right]. 
\end{align}
In comparison with the standard case, we see that the expression of
$\epsilon _1$ contains a $\gamma$ factor in the denominator. This merely
expresses the fact that, even if the potential is not flat, inflation
may occur provided $\gamma \gg 1$. Let us also recall that the above
definition is fact valid for any k-inflation model with the replacement
$\gamma =1/\cs$.

\par

In fact, one does not need more to derive the  slow-roll trajectory
for any DBI model. Let us first notice that Eq.~(\ref{eq:gamma2})
can be recast into
\begin{equation}
\label{eq:trajN2}
\left(\dfrac{\dd N}{\dd\varphi}\right)^2 =
\dfrac{\gamma^{2}H^{2}}{(\gamma^{2}-1)T}\,.
\end{equation}
The Lorentz factor $\gamma$ can now be expressed exclusively in terms of
$H$ and $\varphi$. Indeed, Eq.~(\ref{eq:friedmanndbi2}) together with
$\dot{H}=\dot{\varphi} \, \ud H/\ud \varphi $ yields
\begin{equation}
\label{eq:phidotdbi} 
\dot{\varphi}=-\frac{2}{\kappa\gamma }\dfrac{\ud H}{\ud\varphi}\, , 
\end{equation} 
which can be used to replace $\dot{\varphi}$ in
Eq.~(\ref{eq:gamma2}). Solving for $\gamma$ leads to
\begin{equation}
\label{eq:gamma3}
\gamma(\varphi)=\sqrt{1+\dfrac{4}{\kappa^{2}T}
\left(\dfrac{\ud H}{\ud \varphi}\right)^2} \, .
\end{equation}
Therefore, despite the fact that $\gamma $ contains a $\dot{\varphi }$
factor, it can be viewed as a function of the inflaton field only.
Moreover, as shown in Ref.~\cite{Lorenz:2007ze},
Eq.~(\ref{eq:friedmanndbi}) can be recast as
\begin{equation}
\label{eq:Hslowroll}
3 H^{2}=\dfrac{\kappa V}{1-
\dfrac{2 \gamma}{3(\gamma+1)}\epsilon_{1}}\,.
\end{equation}
Let us notice that, up to this point, all equations are exact. To
proceed further, we use the slow-roll approximation to simplify
Eq.~(\ref{eq:Hslowroll}) by assuming $\epsilon_1 \ll 1$. In this
limit, since $\gamma \ge 1$, one has
\begin{equation}
\label{eq:approxH}
H^2 \simeq \dfrac{1}{3} \kappa V.
\end{equation}
The DBI analogue to the standard slow-roll trajectory is readily
obtained by replacing $\gamma$ in Eq.~(\ref{eq:trajN2}) from its
expression~(\ref{eq:gamma3}) and using Eq.~(\ref{eq:approxH}) for the
Hubble parameter,
\begin{equation}
\label{eq:trajectory2}
N(\varphi) = \mp \kappa\int ^{\varphi}_{\varphi_{\rm ini}}
\sqrt{\left(\dfrac{V}{\deriv{V}{\psi}}\right)^{2}
+\dfrac{1}{3} \dfrac{V}
{\kappa T}} \, \dd \psi .
\end{equation}
As a result, only the knowledge of $V$ and $T$ is required to
calculate the DBI slow-roll trajectory, just as knowing the potential
accomplishes the same goal in the standard case. Let us mention again
that the only assumption that goes into obtaining
Eq.~(\ref{eq:trajectory2}) is $\epsilon_{1}\ll 1$. As we will see in
the next section, additional approximations are nevertheless required
at the perturbative level.

\par

The new degrees of freedom introduced by the warp function $T(\varphi)$
suggest to define an additional hierarchy of
parameters~\cite{Chen:2006nt, Bean:2008ga}. In fact, in the same way
that the $\epsilon_i$ encode the Hubble parameter evolution, it is
convenient to consider their equivalent in terms of the ``sound
horizon''. Therefore, we define the $\delta_{i}$, the sound flow
functions, in a way similar to the Hubble flow parameters, but
starting with the sound speed $\cs$:
\begin{equation}
\label{eq:defdels}
\delta_{n+1}=\frac{\ud \ln \vert \delta_n \vert}{\ud N}\,, \qquad
\delta_{0} \equiv \dfrac{{\cs}_{\mathrm{in}}}{\cs}\,.
\end{equation} 
In the case of DBI inflation, one gets for the two first parameters
\begin{align}
\delta_{1}&= -\frac{2}{\kappa\gamma}\,\frac{1}{\gamma}
\frac{{\rm d}\gamma}{{\rm d}\varphi}\frac{1}{H}
\frac{{\rm d}H}{{\rm d}\varphi}\, ,\\
\delta_{2}&=\frac{2}{\kappa\gamma}\Biggl[\frac{2}{\gamma}
\frac{{\rm d}\gamma}{{\rm d}\varphi}\frac{1}{H}
\frac{{\rm d}H}{{\rm d}\varphi}-\frac{{\rm d}^{2}
\gamma/{\rm d}\varphi^{2}}{{\rm d}\gamma/{\rm d}\varphi}
\frac{1}{H}\frac{{\rm d}H}{{\rm d}\varphi}+\frac{1}{H^{2}}
\left(\frac{{\rm d}H}{{\rm d}\varphi}\right)^{2}
\nonumber \\
& -\frac{1}{H}
\frac{{\rm d}^{2}H}{{\rm d}\varphi^{2}}\Biggr] .
\end{align}
The full hierarchy is therefore given by the combined set
$\left(\epsilon_{i},\delta_{i}\right)$. Let us notice that, as in
standard inflation, it is also possible to introduce various sets of
flow parameters and one could also have a set of potential and warp
function based parameters. Such an alternative hierarchy
$\left(\epsilon_{_{\rm V}},\epsilon_{_{\rm T}}\right)$ is summarized
in the Appendix~\ref{app:srparam} for completeness. Finally, the
$\epsilon_{i}$ and $\delta_{i}$ defined here correspond to a subset of
the ``brane inflation flow functions'' previously defined in
Refs.~\cite{Shandera:2006ax,Bean:2007hc,Kinney:2007ag}.

\subsection{K-inflationary perturbations}
\label{sec:dbipert}

In this subsection, we now turn to the theory of cosmological
perturbations in k-inflation.

\subsubsection{Equation of motion for the scalar modes}
\label{sec:eomscalar}

The main gauge-invariant equations for the cosmological perturbations
are reviewed in Appendix~\ref{app:pert}. There, it was shown that the
(Fourier amplitude of the) Mukhanov-Sasaki variable obeys the
following equation~\cite{Garriga:1999vw}
\begin{equation}
\label{eq:eomv}
v_{\bf k}''+\left(\cs^2k^{2}
-\frac{z''}{z}\right)v_{\bf k}=0 ,
\end{equation}
where a prime denotes a derivative with respect to conformal time and
where the function $z$ is given by the following expression:
\begin{equation}\label{eq:z}
z\equiv \frac{a\varphi'}{\calH}\left(\frac{1}{\cs}\right)^{3/2}
=\sqrt{\frac{2}{\kappa}}\frac{a}{\cs}\sqrt{\epsilon _1}\,.
\end{equation}
As in the standard case, one obtains the equation of a parametric
oscillator. However, there is an important twist: the sound speed
$\cs$ is no longer equal to unity but is now a time-dependent quantity
which prevents the use of standard techniques to find solutions of
this equation. In the DBI case, it is given by $\cs=1/\gamma \le
1$ and can take very low values.

\par

As in the standard case, one can entirely express the effective
potential in Eq.~(\ref{eq:eomv}) in terms of the Hubble and sound flow
functions,
\begin{align}
\label{eq:U}
\dfrac{\left(a\sqrt{\epsilon_1}/\cs\right)''}
{\left(a\sqrt{\epsilon_1}/\cs\right)}
&=\calH^{2}\biggl[2-\epsilon_{1}+\frac{3}{2}\epsilon_{2}
+\frac{1}{4}\epsilon_{2}^{2}-\frac{1}{2}\epsilon_{1}\epsilon_{2}
+\frac{1}{2}\epsilon_{2}\epsilon_{3}
\nonumber \\
& +(3-\epsilon_1+\epsilon_2)\delta_1
+\delta_1^2+\delta_{1}\delta_2\biggr] .
\end{align}
As expected, there are additional terms proportional to the sound flow
parameters.

\par

As usual, the final goal is to compute the two-point correlation
function or, in Fourier space, the power spectrum. For the scalar
modes, its expression reads
\begin{equation}
\label{eq:defP}
{\cal P}_{\zeta}\equiv \frac{k^3}{2\pi ^2}
\left\vert \zeta _{\bf k}\right\vert ^2
=
\frac{k^3}{4\pi ^2}
\frac{\cs^2\kappa\left\vert v_{\bf k}\right\vert
^2}{a^2\epsilon _1}\, ,
\end{equation}
where $\zeta_{\bf k}=v_{\bf k}/z$ is the comoving curvature
perturbation. In order to estimate this quantity, one has to integrate
the equation of motion~(\ref{eq:eomv}) from a set of initial
conditions over the background solution,  given by
Eq.~(\ref{eq:trajectory2}) for the DBI models. We now turn to the
question of the initial conditions.

\subsubsection{Initial conditions}

The effective time-dependent frequency of Eq.~(\ref{eq:eomv}) is given
by
\begin{equation}
\label{eq:omegasqrd}
\omega^{2}(k,\eta)=\cs^2k^2
-\dfrac{\left(a\sqrt{\epsilon_1}/\cs\right)''}
{\left(a\sqrt{\epsilon_1}/\cs\right)}\, .
\end{equation}
A well-defined and well-motivated initial state can be chosen in the
adiabatic regime for which a WKB solution exists, \ie for
\begin{equation}
\label{eq:wkbcondition}
\left \vert \frac{Q}{\omega ^2}\right \vert \ll 1\, ,
\end{equation}
where the quantity $Q$ is
\begin{equation}
\label{eq:Qwkb}
Q(k,\eta)\equiv \frac{3}{4}\frac{\omega'^{2}}{\omega^{2}}
-\frac{\omega''}{2\omega}\, .
\end{equation}
In the standard case, the modes of astrophysical interest today are, at
the beginning of inflation, such that their wavelength is smaller than
the Hubble radius. This implies that $\omega \sim k$ and the quantity
$Q$ vanishes. As a consequence, the condition $\vert Q/\omega ^2\vert
\ll 1$ is obviously satisfied and the WKB state
\begin{align}
v_{\bf k}(\eta ) =\dfrac{\exp \displaystyle \left[i\int^{\eta } 
\omega (k,\tau ){\rm d}\tau \right]}{\sqrt{2\omega(k,\eta)}} 
\simeq \frac{1}{\sqrt{2k}}
{\rm e}^{ik\left(\eta-\eta _{\rm ini}\right)} \, ,
\end{align}
is the preferred initial state.

\par

\begin{figure}
\begin{center}
\includegraphics[width=0.48\textwidth]{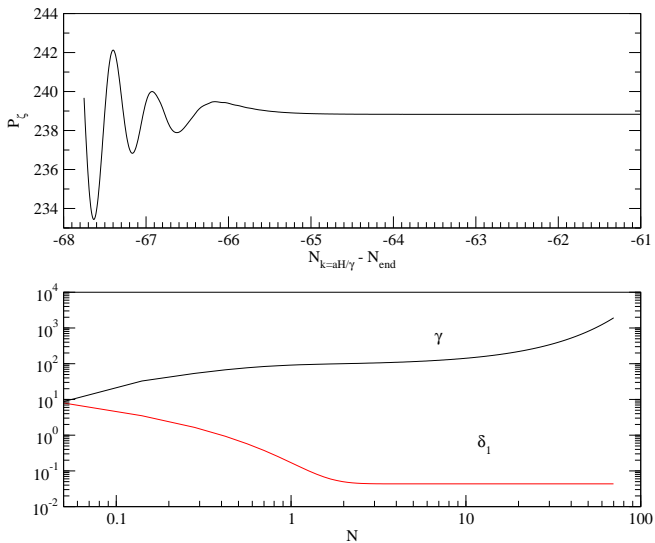}
\caption{Scalar power spectrum as a function of the number of e-folds
  $N_k$ at which a given mode $k$ crossed the sound horizon in the
  case of the chaotic Klebanov--Strassler model (see
  Sect.~\ref{sec:chaotic}). The oscillations on large scales come from
  the violation of the WKB condition while the initial state is
  enforced to be a Bunch-Davies vacuum. The bottom panel shows the
  sound flow parameters as a function of the number of e-folds:
  $\delta_1\gg 1$ initially triggers the WKB violation.}
\label{fig:wkbviolation}
\end{center}
\end{figure}

However, in the k-inflationary case, the time-dependence of the sound
speed brings new complications into this
question~\cite{Garriga:1999vw,Alishahiha:2004eh,Bean:2007eh,
Kinney:2007ag}. Indeed, even if initially, the modes are within the
sound horizon and such that $\omega \sim \cs k$, the effective frequency
is still a time-dependent quantity and, therefore, it is \emph{a priori}
not obvious that a well-defined state can be chosen in this context. One
has to check for each mode that the quantity $Q/\omega ^2$ is indeed
small. From Eqs.~(\ref{eq:omegasqrd}) and (\ref{eq:Qwkb}), this one can
be expressed in terms of the sound flow functions and reads
\begin{equation}
\frac{Q}{\omega ^2}=\frac{a^2H^2}{2\cs^2k^2}\left(\delta _1
-\epsilon _1 \delta _1+\delta _1\delta _2+\frac12 \delta _1^2\right).
\end{equation}
Assuming the slow-roll conditions  on the $\epsilon_i$  are
satisfied, the WKB condition (\ref{eq:wkbcondition}) can still be
violated as soon as $\cs k/\calH \simeq \delta_1$, and thus even for
the modes deep inside the sound horizon provided $\delta_1$ is big
enough.  As an example of such a situation,
Fig.~\ref{fig:wkbviolation} shows the scalar power spectrum obtained
from a numerical integration of Eq.~(\ref{eq:eomv}) in a case where
$\delta_1 \gg 1$ initially, the warp function and potential being
those of the chaotic Klebanov--Strassler model, see
Sect.~\ref{sec:chaotic}. Enforcing the modes to start in the
Bunch--Davies vacuum is no longer justified and leads to oscillations
in the power spectrum. Clearly, such a situation essentially concerns
the modes crossing the sound horizon soon after the beginning of
inflation, the ones for which $\cs k/\calH $ cannot be chosen big
enough to satisfy Eq.~(\ref{eq:wkbcondition}).

\par

The conclusion is that, in order to be able to choose a well-motivated
initial state for a given Fourier mode in k-inflation, it is not
enough to have a wavelength smaller than the sound
horizon. Additional conditions on the sound flow parameters
$\delta _1$ and $\delta _2$ are also necessary. As shown before, a
situation where $\lambda_k=2\pi a/k $ is smaller than the sound
horizon and $Q/\omega ^2\gg 1 $ can easily been designed. In such a
case, our ability to work with a well-defined initial state is lost.

\subsection{The uniform approximation}
\label{sec:uniformapprox}

At first order in the Hubble and sound flow functions, one can use the
uniform approximation to solve the mode equation. The uniform
approximation was developed in Refs.~\cite{Habib:2002yi,
  Habib:2004kc}. For this purpose, it is convenient to re-write
Eq.~(\ref{eq:eomv}) as
\begin{equation}
v_{\bf k}''+\left(\cs^2k^2
-\frac{\nu ^2-1/4}{\eta ^2}\right)v_{\bf k}=0 , 
\end{equation}
where the function $\nu (\eta )$ can be calculated from the effective
potential given in Eq.~(\ref{eq:U}). Following
Refs.~\cite{Habib:2002yi,Habib:2004kc}, one also defines the two
following functions:
\begin{equation}
g(\eta )\equiv \frac{\nu ^2}{\eta ^2}-\cs^2k^2\, ,\quad
q(\eta )\equiv -\frac{1}{4\eta ^2}\, .
\end{equation}
The so-called turning point is defined by the condition $g(\eta _*)=0$
and, for each mode, occurs at the time $\eta_*(k)$ such that
\begin{equation}
\label{eq:turningpoint}
  k \eta_*=-\frac{\nu_*}{\cs{}_{*}}\, .
\end{equation}
The uniform approximation tells us that the Mukhanov variable $v_{\bf
  k}$ can be expressed as~\cite{Habib:2002yi,Habib:2004kc}
\begin{equation}
\label{eq:uniformsol}
v_{\bf k}(\eta) = A_{\bf k}\left(\frac{f}{g}\right)^{1/4}\Airy{f}
+B_{\bf k}\left(\frac{f}{g}\right)^{1/4}\Biry{f} \, ,
\end{equation}
where $A_{\bf k}$ and $B_{\bf k}$ are two constants to be determined
from the initial conditions and $\Airy{x}$ and $\Biry{x}$ denote the
Airy functions of first and second kind respectively. The function
$f(k,\eta)$ is defined by
\begin{equation}
\label{eq:unidef}
f(k,\eta) = \dfrac{\left\vert\eta-\etaS\right\vert}{\eta-\etaS} 
\left\vert \dfrac{3}{2}
\int_{\etaS}^\eta \ud \tau \sqrt{\left\vert g(\tau)\right\vert} 
\right\vert^{2/3} .
\end{equation}

As already discussed above, we assume that adiabaticity is valid
initially, then choosing the initial conditions as an initial state of
the WKB form yields
\begin{equation}
A_{\bf k}=iB_{\bf k}\, ,\quad B_{\bf k}=\sqrt{\frac{\pi}{2}}{\rm
 e}^{i\theta} ,
\end{equation}
where $\theta $ is just an unimportant scale-independent phase factor
that will be ignored in the following. The
solution~(\ref{eq:uniformsol}) is now completely specified.
 
\par

After the turning point, using the asymptotic behavior of the Airy
functions, the solution can be written as
\begin{equation}
\label{eq:unisuper}
v_{\bf k}(\eta )\simeq \frac{B_{\bf k}}{g^{1/4}\pi^{1/2}}
\exp\left(\dfrac{2}{3} f^{3/2} \right) , 
\end{equation}
and one still has to calculate the integral in
Eq.~(\ref{eq:unidef}). For $\eta>\etaS$, it simplifies to
\begin{equation}
\label{eq:fsup}
\dfrac{2}{3} f^{3/2}(k,\eta) = \int _{\eta _*}^{\eta }{\rm d}
\tau \sqrt{\frac{\nu ^2}{\tau
 ^2}-\cs^2k^2}\, .
\end{equation}
The functions $\nu (\eta )$ and $1/\cs(\eta)$ can be expanded in terms
of the Hubble and sound flow functions. For instance, assuming that
$1/\cs$ admits a polynomial expansion around $\etaS$, one has 
\begin{equation}
\label{eq:gammaexpand}
\frac{1}{\cs(\eta)} = \sum_{n=0}^{\infty} 
 \left(\frac{1}{\cs}\right)^{(n)}_{\eta _*}
\frac{(\eta-\etaS)^n}{n!}\, .
\end{equation}
At first order, one finds from Eqs.~(\ref{eq:defeps})
and~(\ref{eq:defdels})
\begin{equation}
\label{eq:gammaderivs}
\left(\frac{1}{\cs}\right)^{(n)}
=\frac{1}{\cs}(n-1)! \, \calH^n + \order{\epsilon
\delta},
\end{equation}
where $\order{\epsilon \delta}$ stands for all terms of order two in the
$\epsilon_{i}$, $\delta_{i}$ or mixed. Let us notice that all terms in
Eq.~(\ref{eq:gammaexpand}) should be considered at first order in
$\epsilon_i$ and $\delta_i$. Plugging Eq.~(\ref{eq:gammaderivs}) into
Eq.~(\ref{eq:gammaexpand}) yields an infinite sum that can however be
resummed into
\begin{equation}
\dfrac{\cs {}_*}{\cs(\eta)} = 1 - \deltaoneS 
\ln\left[1-\calHS(\eta-\etaS)\right],
\end{equation}
a star indicating that the corresponding quantity is evaluated at the
turning point defined above. For consistency, $\calHS$ has still to be
expanded. Using
\begin{equation}
\label{eq:hubbleexp}
 \calH(\eta) = -\dfrac{1+\epsilon_1}{\eta} 
+ \order{\epsilon \delta},
\end{equation} 
one finally gets
\begin{equation}
\label{eq:gammafinal}
\frac{1}{\cs(\eta)} = \frac{1}{\cs {}_*}
\left(1 - \deltaoneS \ln\dfrac{\eta}{\etaS}
\right) + \order{\epsilon \delta} .
\end{equation}
Along the same line of reasoning, one would show that the function
$\nu(\eta)$ reads
\begin{equation}
\label{eq:nufinal}
\nu^2(\eta)=\dfrac{9}{4} + 3 \epsoneS + \dfrac{3}{2} \epstwoS + 3
\deltaoneS + \order{\epsilon \delta} = \nu_*^2 + \order{\epsilon \delta}.
\end{equation}
One can check that this expression matches with the standard result by
setting $\deltaoneS=0$ (see Ref.~\cite{Stewart:1993bc}). Let us notice
that one may avoid the infinite re-summation by performing an expansion
directly in terms of the number of e-folds. Indeed, a Taylor expansion
of $1/\cs(N)$ is also a flow expansion. For instance, one has
\begin{equation}
\label{eq:gammaefold}
\frac{1}{\cs(N)}= \frac{1}{\cs {}_*}+\frac{1}{\cs {}_*} 
\deltaoneS \left(N-N_*\right) + \order{\delta^2}.
\end{equation}
To recover the conformal time dependency one may use
Eq.~(\ref{eq:hubbleexp}) to get
\begin{equation}
\label{eq:efoldtoeta}
  N-N_* = \ln\dfrac{a}{a_*} \simeq \ln
  \dfrac{(-\eta)^{-1-\epsilon_1}}{(-\etaS)^{-1-\epsoneS}} =
  \ln\dfrac{\etaS}{\eta} + \order{\epsilon}.
\end{equation}
Plugging the previous equation into Eq.~(\ref{eq:gammaefold})
immediately gives Eq.~(\ref{eq:gammafinal}).

\par

Let us now return to the calculation of the function $f(k,\eta)$.
Inserting Eqs.~(\ref{eq:gammafinal}) and (\ref{eq:nufinal}) into the
integral~(\ref{eq:fsup}), and defining the new variable
\begin{equation}
\label{eq:defuketa}
w \equiv \dfrac{\cs {}_*k\eta }{\nu _*}\, ,
\end{equation}
one obtains at first order in the  Hubble and sound flow
parameters
\begin{align}
\label{eq:fequal}
\dfrac{2}{3} f^{3/2}(w) & = \left(-\dfrac{3}{2} - \epsoneS -
  \dfrac{1}{2} \epstwoS + \dfrac{1}{2} \deltaoneS \right) \nonumber \\
& \times \left(
  \sqrt{1-w^2} + \ln \dfrac{-w}{1+\sqrt{1-w^2}} \right) \nonumber \\ 
& -\dfrac{3}{2} \deltaoneS \sqrt{1-w^2} \ln(-w) .
\end{align}
where we have used Eq.~(\ref{eq:turningpoint}). This expression will
be used in the following to  derive the power spectra.

\subsection{Power spectra and spectral index}
\label{sec:powerspectra}

\subsubsection{Scalar power spectrum}
\label{sec:index}

We are now in a position to estimate the power spectrum of scalar
perturbations. The comoving curvature perturbation being constant on
super-sonic length scales, see Appendix~\ref{app:pert}, the power
spectrum is obtained from Eq.~(\ref{eq:defP}), using
Eqs.~(\ref{eq:unisuper}) and (\ref{eq:fequal}) in the limit $w
\rightarrow 0$. Lengthy but straightforward calculations give, at
first order in the Hubble and sound flow functions,
\begin{align}
\label{eq:zetapower}
{\cal P}_{\zeta} &= \dfrac{H_*^2}{\pi \mpl^2\epsoneS \cs {}_*}
\left(18 \ue^{-3}\right) \Biggl[1
  - 2\left(\dfrac{4}{3}-\ln 2 \right) \epsoneS 
\nonumber \\ &  - 
  \left(\dfrac{1}{3} - \ln 2 \right) \epstwoS + 
  \left(\dfrac{7}{3} - \ln 2\right) \deltaoneS \Biggr].
\end{align}
Notice that the above expression is still an implicit function of $k$
through its dependence on $\etaS$. Let us also remark the presence of
the factor $18 \ue^{-3}\sim 0.896$ in the overall amplitude, which is
typical for the WKB and uniform approximations. As discussed in
Ref.~\cite{Martin:2002vn}, this is rather unfortunate since this factor
damages the approximation of the overall amplitude down to the $10\% $
level. However, as shown for instance in Ref.~\cite{Casadio:2004ru},
this problem can be rather easily fixed. Very roughly, one can
renormalize $18 \ue^{-3}$ to one to recover the exact
amplitude~\cite{Martin:2002vn}. The spectral index is, on the contrary,
predicted accurately by the WKB and uniform approximations.

\par

To remove the implicit dependence in $k$ hidden in $\etaS$, we define
a pivot wavenumber $k_\piv$ and expand all terms around an unique
conformal time $\etaP$ which is the time when the pivot scale
crossed the ``sound horizon'', namely
\begin{equation}
\label{eq:etapiv}
-k_\piv \etaP = \frac{1}{\csP}\, .
\end{equation}
Then, one can use the flow expansions for $\calH$, $1/\cs$, and the
$\epsilon_i$, $\delta_i$, but now around $\eta_\piv$. In this case,
Eq.~(\ref{eq:zetapower}) becomes
\begin{eqnarray}
\label{eq:scalarspectrum}
{\cal P}_{\zeta} &=& \dfrac{H_\piv^2}{\pi \mpl^2\epsoneP\csP}
\left(18 \ue^{-3}\right)
 \Biggl[1
    - 2(D+1) \epsoneP - 
    D \epstwoP \nonumber \\ & & + (D+2) \deltaoneP - \left(2\epsoneP +
     \epstwoP -\deltaoneP \right) \ln
    \dfrac{k}{k_\piv} \Biggr] ,
\end{eqnarray}
where we have defined $D\equiv 1/3 - \ln 3$. From the above expression,
one can already read off the spectral index of scalar perturbations,
\begin{equation}
\nS-1=-2\epsoneP - \epstwoP +\deltaoneP\, .
\end{equation}
The standard expression is corrected by a term, $\deltaoneP$, which
takes into account the time-dependence of the sound speed.

\subsubsection{Tensor power spectrum}
\label{sec:tensor}

Although the equations of motion and evolution of the tensor
perturbations are not affected by a varying sound speed in the scalar
sector, there is however a subtle effect associated with the choice of
the e-fold at which one evaluates the Hubble and sound flow functions
entering the Taylor expansion of the power spectrum.

\par

As shown in Ref.~\cite{Martin:2002vn}, around the pivot scale $k_\piv$,
the tensor power spectrum in the WKB, or uniform approximation, at first
order in slow-roll, reads
\begin{equation}
\label{eq:phstd}
\calP_h(k) = \dfrac{16 H_{\pivhub}^2}{\pi \mpl^2} \left(18
  \ue^{-3}\right) \left[1 - 2(D+1) \epsonePH - 2 \epsonePH \ln
  \dfrac{k}{k_\piv} \right], 
\end{equation}
where all background quantities are evaluated at the time
$\etaPH$ such that
\begin{equation}
\label{eq:etahub}
-k_\piv \etaPH = 1,
\end{equation}
which is obviously different than $\etaP$ in Eq.~(\ref{eq:etapiv}). It
is therefore convenient to express $H_{\pivhub}$ and $\epsonePH$ in
terms of the parameters evaluated at $\eta=\etaP$. Using
Eq.~(\ref{eq:efoldtoeta}) with $\etaPH = \csP\etaP$, at first order in
the sound flow parameters, one gets
\begin{align}
\label{eq:tensorspectrum}
\calP_h(k) & = \dfrac{16 H_\piv^2}{\pi \mpl^2} \left(18
  \ue^{-3}\right) \Biggl[1 - 2\left(D + 1 + \ln \frac{1}{\csP}\right)
\epsoneP \nonumber \\ & - 2 \epsoneP \ln \dfrac{k}{k_\piv} \Biggr].
\end{align}
Let us remark that, in the amplitude of the power spectrum, the
coefficient in front of the first slow-roll parameter is different as
compared to the standard case. Indeed, in the standard case, $\csP=1$
and $\ln (1/\csP)$ vanishes.

\par

Eqs.~(\ref{eq:scalarspectrum}) and~(\ref{eq:tensorspectrum})
constitute one of the main result of this article. They represent the
general scalar and tensor power spectra of k-inflation, valid at first
order in the Hubble and sound flow parameters. In
Ref.~\cite{Lorenz:2008je}, they have recently been compared to the WMAP5
data~\cite{Gold:2008kp, Hill:2008hx, Hinshaw:2008kr, Nolta:2008ih,
Dunkley:2008ie, Komatsu:2008hk}.

\subsection{Running of the spectral index and higher order corrections}
\label{sec:running}

Having obtained the scalar perturbation power spectrum from the uniform
approximation, we will now put this result to use in calculating the
spectral index $\nS$ and the running $\alpha_{\rm s}$ of the
spectral index using the method of Ref.~\cite{Schwarz:2004tz}. Expanding
the scalar perturbation power spectrum around the pivot scale $k_\piv$
in terms of $\ln(k/k_\piv)$, one has
\begin{equation}
\label{eq:exp}
\mathcal{P}_\zeta(k)=\tilde{\mathcal{P}}_\zeta(k_\piv) \sum_{n\geq0}
\frac{a_{n}}{n!}\,\ln^{n}\left(\frac{k}{k_\piv}\right).
\end{equation}
At zeroth order, the spectrum
is given by the leading term of Eq.~(\ref{eq:scalarspectrum}), namely
\begin{equation}
\label{eq:spec0}
\tilde{\mathcal{P}}_\zeta(k_\piv)=\frac{H_\piv^{2}}
{\pi\mpl^{2}\epsoneP\csP}\,\left(18 \ue^{-3}\right).
\end{equation}
Since the physical power spectrum must not depend on the pivot scale,
$\dd \mathcal{P}_\zeta(k)/\dd\ln k_\piv=0$, one may establish the
recursion relation~\cite{Schwarz:2004tz}
\begin{align}
\label{eq:recursion}
 a_{n+1}&= \dfrac{\dd\ln \tilde{\mathcal{P}}_\zeta}{\dd\ln k_\piv}
 a_{n}+\dfrac{\dd a_{n}}{\dd\ln k_\piv} \nonumber \\
  &= \dfrac{1}{1-\epsoneP+\deltaoneP} \left(\dfrac{\dd a_{n}}{\dd
      N_\piv} + \dfrac{\dd \ln\tilde{\mathcal{P}}_\zeta} {\dd N_\piv}
    a_{n} \right),\quad n\geq0,
\end{align}
where in the last expression we have used
\begin{equation}
\dd\ln k_{\piv}=(1-\epsoneP+\deltaoneP)\dd N_\piv .
\end{equation}
Using Eq.~(\ref{eq:spec0}), one gets
\begin{equation}
\dfrac{\dd \ln\tilde{\mathcal{P}}_\zeta}{\dd
  N_\piv}=-2\epsoneP - \epstwoP + \deltaoneP .
\end{equation}
In terms of the expansion (\ref{eq:exp}), the first coefficient $a_{0}$
is determined by the spectral amplitude, while $a_{1}$ is related to the
spectral index, and $a_{2}$ to the running. Note that if we know $a_{0}$
to $q$th order in $\left(\epsilon_{i},\delta_{i}\right)$, we can
determine $a_{n}$ to the order $q+n$. From the uniform approximation
[see Eq.~(\ref{eq:scalarspectrum})] we know that
\begin{equation}
\label{eq:a0}
a_{0}=1-2(D+1)\epsoneP - D\epstwoP + (D+2)\deltaoneP,
\end{equation}
at first order. The recurrence relation, up to second order, therefore
gives
\begin{eqnarray}
a_{1} &=&-2\epsoneP-\epstwoP+\deltaoneP
+2(2D+1)\,\epsoneP^{2} \nonumber \\ & &
-(4D+3)\,\epsoneP\deltaoneP
+(2D-1) \epsoneP
\epstwoP 
+D \epstwoP^{2}
\nonumber \\ & &
-D \epstwoP\epsthreeP-(2D+1) \epstwoP \deltaoneP
+(D+1) \deltaoneP^{2}\nonumber \\ & & +(D+2) \deltaoneP \deltatwoP .
\end{eqnarray}
One more iteration of Eq.~(\ref{eq:recursion}) allows to determine
$a_{2}$ up to third order,
\begin{eqnarray}
a_{2} & = & 4 \epsoneP^{2} + 2 \epsoneP \epstwoP - 4\epsoneP
\deltaoneP+ \epstwoP^{2} - 2 \epstwoP \deltaoneP - \epstwoP
\epsthreeP \nonumber \\ & &
+\deltaoneP^{2} + \deltaoneP \deltatwoP +
6 \epsoneP^{2} \epstwoP + 12D \epsoneP^{2} \deltaoneP  
\nonumber \\ & &
-(2D-1) \epsoneP\epstwoP^{2}+3(2D-1) \epsoneP \epstwoP
\deltaoneP-6D \epsoneP \deltaoneP^{2} 
\nonumber \\ & &
+3D\epstwoP^{2} \deltaoneP -3D \epstwoP \deltaoneP^{2} +D
  \deltaoneP^{3} -3D \epstwoP \epsthreeP \deltaoneP
\nonumber \\ & &  
+3(D+1) \deltaoneP^{2} \deltatwoP
-8D \epsoneP^{3}+2(2D-1) \epsoneP \epstwoP \epsthreeP 
\nonumber \\ & &
- 6(D+1) \epsoneP \deltaoneP
\deltatwoP -D \epstwoP^{3}+3D \epstwoP^{2} \epsthreeP 
\nonumber \\ & &
-3(D+1) \epstwoP \deltaoneP \deltatwoP
- D\epstwoP \epsthreeP^{2} - D \epstwoP \epsthreeP
\epsfourP  \nonumber \\ & &
+(D+2) \deltaoneP \deltatwoP^{2} + (D+2) \deltaoneP
\deltatwoP \deltathreeP.
\end{eqnarray}
The relation between the $a_{i}$ and $n_{\rm s}$ is obtained using the
definition
\begin{equation}
\nS-1 = \left(\frac{\dd\ln\mathcal{P}}{\dd\ln k}\right)
_{k=k_\piv} ,
\end{equation}
which, when compared to the corresponding derivative of
Eq.~(\ref{eq:exp}), leads to
\begin{equation}
\label{eq:ns_coeff}
\nS-1= \dfrac{a_{1}}{a_{0}}\,,
\end{equation}
and at second order in the Hubble and sound flow parameters, one gets
\begin{eqnarray}
\label{eq:ns}
\nS-1 & = & -2\epsoneP-\epstwoP+\deltaoneP-2\epsoneP^{2} 
- (2D+3) \epsoneP\epstwoP\nonumber \\ & & +3\epsoneP\deltaoneP +
\epstwoP\deltaoneP - D \epstwoP\epsthreeP-\deltaoneP^{2} 
\nonumber \\ & & + (D+2)
\deltaoneP\deltatwoP \, .
\end{eqnarray}
The same calculation can be repeated for the running. Its definition
reads
\begin{equation}
  \alphaS = \left(\frac{\dd^{2}\ln\mathcal{P}_\zeta}{\dd\ln^{2}
      k}\right)_{k=k_\piv}\, ,
\end{equation}
and it can be identified with
\begin{equation}
\label{eq:alpha_coeff}
\alphaS = \dfrac{a_{2}}{a_{0}} - \dfrac{a_{1}^{2}}{a_{0}^{2}}\,.
\end{equation}
Up to second order, one finds
\begin{align}
\label{eq:alphaSsecond}
\alphaS & = -2\epsoneP \epstwoP - \epstwoP \epsthreeP +
\deltaoneP \deltatwoP \, , 
\end{align}
We see that the parameter $\delta _2$ appears in the above expression.
Using the tools developed previously, one can even estimate the running
at third order. The result reads
\begin{eqnarray}
\label{eq:alphaS}
\alphaS & = & -2\epsoneP \epstwoP - \epstwoP \epsthreeP +
\deltaoneP \deltatwoP - 6\epsoneP^{2} \epstwoP
+5\epsoneP\epstwoP\deltaoneP 
\nonumber \\ & &
+4\epsoneP\deltaoneP\deltatwoP 
-(2D+3)\epsoneP\epstwoP^{2}-2(D+2)\epsoneP\epstwoP\epsthreeP
\nonumber \\ & &
-D\epstwoP\epsthreeP^{2} -D\epstwoP\epsthreeP\epsfourP
+2\epstwoP\epsthreeP\deltaoneP+\epstwoP\deltaoneP\deltatwoP 
\nonumber \\ & &
-3\deltaoneP^{2}\deltatwoP+(D+2)\deltaoneP\deltatwoP^{2} 
+(D+2)\deltaoneP\deltatwoP\deltathreeP\, .
\end{eqnarray}
{}Finally, from the power spectrum of the tensor perturbations, the
tensor to scalar ratio at first order in the Hubble and sound flow
parameters reads
\begin{equation}
\label{eq:tenstoscal}
r = \dfrac{\calP_h}{\calP_\zeta} = 16 \csP\epsoneP\,.
\end{equation}
We recover that since $\cs \le 1$, $r$ is reduced compared to the
canonical single field dynamics.

\par 

We can compare our approach to the existing results in the literature,
and, at first order, our results agree with those of
Refs.~\cite{Shandera:2006ax, Bean:2007hc, Peiris:2007gz,
  Kinney:2007ag}. However, we would like to stress that at higher
order, and in particular for the running, our results match only with
those of Ref.~\cite{Kinney:2007ag}. Indeed, in
Refs.~\cite{Shandera:2006ax, Bean:2007hc, Peiris:2007gz},
Eq.~(\ref{eq:eomv}) is solved in terms of Bessel functions along the
lines of the standard formalism which assumes that $\cs$ is a
constant. This ends up being an acceptable assumption at zeroth order
only as is clear from Eq.~(\ref{eq:gammaefold}). The spectral index at
first order being, roughly speaking, the derivative of the power
spectrum amplitude at zeroth order, one may indeed obtain its correct
first order expression with the assumption that $\cs$ is
constant. However, if one wants to derive its quadratic corrections,
or the running, then it is necessary to know the correct amplitude of
the power spectra at first order. In Ref.~\cite{Kinney:2007ag},
this goal was achieved using a conveniently chosen transformation of
the time variable to absorb the time dependence of the sound speed,
the resulting equation being integrable at first order.

This argument is at the heart of the present article. If the sound
speed of the perturbations is a time-dependent quantity and if the
first order expression~(\ref{eq:gammaefold}) of $\cs$ is inserted into
Eq.~(\ref{eq:eomv}), then the solution cannot be found by the usual
technique. Therefore, the standard approach cannot be used and this
prompts the use of a different method. This is what was done in
Ref.~\cite{Kinney:2007ag}, using a new time variable, and what is done
in the present paper under the WKB/uniform approximation.

\subsection{Model classification}
\label{sec:classification}

To conclude this section, we generalize the classification of
inflationary models of Ref.~\cite{Schwarz:2004tz} to the DBI case. The
energy density of a DBI inflaton field of Eq.~(\ref{eq:defrhop}) can be
re-written as
\begin{equation}
\label{eq:rho}
\rho=\frac{\gamma^{2}}{\gamma+1}\,\dot{\varphi}^{2}+V(\varphi),
\end{equation}
from which it is easy to see that we recover the standard expression
in the limit $\gamma \rightarrow 1$. Let us refer to the first term in
Eq.~(\ref{eq:rho}) as the kinetic energy contribution, while the second
term represents the potential energy. The Hubble flow functions
$\epsilon_{1}$ and $\epsilon_{2}$ then can be used to study the
respective evolution of these contributions. With
\begin{align}
  \epsilon_{1} &= \dfrac{3\gamma \dot{\varphi}^{2}}{2\rho}\, ,\\
  \epsilon_{2} &= 2\left( \dfrac{\ddot{\varphi}} {H\dot{\varphi}} +
    \epsilon_{1} + \frac{1}{2} \delta_{1} \right) ,
\end{align}
which represents the generalization of Eqs.~(6) and~(7) of
Ref.~\cite{Schwarz:2004tz}, the change in the potential energy is given
by
\begin{equation}
\label{eq:changeV}
\dot{V} = -H\gamma \dot{\varphi}^{2} \left[3+ \frac{\gamma}
  {\gamma+1} (\epsilon_{2}-2\epsilon_{1}) + \frac{\gamma}
  {(\gamma+1)^{2}} \delta_{1}\right].
\end{equation}
Note that again the standard expressions are recovered for
$\gamma\rightarrow1$ since, in this case, also
$\delta_{1}\rightarrow0$. From Eq.~(\ref{eq:changeV}), we see that the
potential energy density can never increase for small values of
$\epsilon_{i}$ and $\delta_{i}$ even if $\gamma $ is large.

\par

The change for the ratio between kinetic and total energy density is
given by 
\begin{equation}
\frac{\dd}{\dd t} \left(\frac{\epsilon_{1}}{3}\right) 
= H \frac{\epsilon_{1}}{3} \epsilon_{2}\, .
\end{equation}
Therefore, $\epsilon_{2}=0$ is the borderline between the regime where
the kinetic energy contribution to $\rho$ increases ($\epsilon_{2}>0$),
and the regime where the kinetic energy contribution decreases
($\epsilon_{2}<0$). For a refined classification of DBI inflationary
models, let us also calculate the time derivative of the kinetic energy
density in Eq.~(\ref{eq:rho}):
\begin{equation}
\frac{\dd}{\dd
  t}\left(\frac{\gamma^{2}}{\gamma+1}\,\dot{\varphi}^{2}\right) =
\frac{\gamma^{2}}{(\gamma+1)^{2}} H\dot{\varphi}^{2} \left[(\gamma+1)
  (\epsilon_{2}-2\epsilon_{1}) +\delta_{1} \right].
\end{equation}
This equation can be viewed as the equivalent of Eq.~(10) of
Ref.~\cite{Schwarz:2004tz}. Hence, the kinetic energy density increases
while
\begin{equation}
\epsilon_{2}>2\epsilon_{1}- \dfrac{\delta_{1}}{\gamma+1}\, ,
\end{equation}
and decreases otherwise. We see that the standard condition,
$\epsilon_2-2\epsilon _1>0$ or $\epsilon_2-2\epsilon _1<0$ is modified
and that the sound flow parameter $\delta _1$ now participates in the
new criterion. This is natural since the factor $\gamma $ appears in the
expression of the kinetic energy. However, in the limit $\gamma
\rightarrow +\infty$, the standard condition is also recovered.

\section{Example models}
\label{sec:examples}

We now illustrate our results in the case of DBI inflation with some
specific choices of $V(\varphi)$ and $T(\varphi)$ considered in the
literature. The DBI slow-roll trajectory permits to calculate the
resulting values of $\gamma$ and the first $\epsilon_{i}$ and
$\delta_{i}$ parameters, and therefore the shape of the scalar
primordial power spectrum through Eq.~(\ref{eq:scalarspectrum}). As a
warm up, we start with power-law inflation, where, as it is the case in
the standard situation, exact solutions are available. Then we turn
to the more important case of brane inflation that we discuss in some
detail.

\subsection{DBI power-law inflation}
\label{sec:dbipowerlaw}

The DBI analogue of power-law inflation has been studied in
Refs.~\cite{Spalinski:2007dv,Spalinski:2007qy}. As we show in the
following, it is particularly convenient for testing the previous
approximations since the perturbation equations are exactly solvable in
this case.

\par

Looking for a power law behavior of the scale factor in terms of the
conformal time, one finds that the warp function and the potential are
given by~\cite{Spalinski:2007dv,Spalinski:2007qy,Spalinski:2007un}
\begin{align}
T(\varphi)&= T_0\exp\left[-\sqrt{\frac{2\gamma \kappa}{p}}
(\varphi-\varphi_0)\right] ,\\
V(\varphi)&= V_0\exp\left[-\sqrt{\frac{2\gamma \kappa}{p}}
(\varphi-\varphi_0)\right],
\end{align}
where the two constants $T_0$ and $V_0$ are related by 
\begin{equation}
V_0=T_0\frac{\gamma -1}{\gamma}\left[\frac{3p}{2}(\gamma +1)-\gamma 
\right].
\end{equation}
The solution of the Einstein equations are such that the Lorentz factor
$\gamma$ is a constant and the scale factor and scalar field can be
expressed as
\begin{eqnarray}
\label{eq:ansatz-a}
a(t) &=& a_{0}\left(\frac{t}{t_0}\right)^{p} , \\
\varphi(t) &=& \varphi_0+\sqrt{\frac{2p}{\kappa \gamma}}
\ln \left(\frac{t}{t_0}\right) .
\end{eqnarray}
In this case, one has also the relation 
\begin{equation}
T_0=\dfrac{2p\gamma}{\kappa t_0^2(\gamma ^2-1)}\,.
\end{equation}

\par

In terms of conformal time $\eta$, the scale factor is also of the
power-law form
\begin{equation}
a(\eta)=\ell _0\left\vert \eta \right \vert
^{1+\beta},
\end{equation}
with
\begin{equation}
\beta =-\frac{2p-1}{p-1}\, .
\end{equation}
One can check that when $p\rightarrow \infty$, $\beta $ goes to $-2$ and
one recovers the de Sitter case.

\par

Let us now turn to the calculation of the perturbations. The equation of
motion (\ref{eq:eomv}) for the Mukhanov-Sasaki variable takes the form
\begin{equation}
v_{\bf k}''+\left[\frac{k^2}{\gamma ^2}-\frac{\beta (\beta
		      +1)}{\eta ^2}\right]v_{\bf k}=0 ,
\end{equation}
where $\gamma$ is now a constant. This equation can be solved
explicitly in terms of Bessel functions and the solution reads
\begin{align}
\label{eq:solvpowerlaw}
v_{\bf k}(\eta)&= \left(k\eta \right)^{1/2}
\Biggl[ A_{\bf k}\besselJ{\beta +1/2}
\left(\dfrac{k\eta }{\gamma}\right)
\nonumber \\  
  &+ B_{\bf k}
\besselJ{-(\beta +1/2)}
\left(\dfrac{k\eta }{\gamma}\right)\Biggr] ,
\end{align}
where $A_{\bf k}$ and $B_{\bf k}$ are two scale-dependent arbitrary
constants. As usual, one requires the initial state, evaluated in the
limit $k\eta /\gamma \rightarrow -\infty$ to be of the WKB type. This
completely fixes $A_{\bf k}$ and $B_{\bf k}$ which are given by
\begin{align}
A_{\bf k} &= -B_{\bf k}\ue^{i\pi (\beta +1/2)}\, ,\\ 
B_{\bf k} &= \sqrt{\frac{\pi}{4k}}\frac{\ue^{i\pi(\beta
+1/2)/2-i\pi/4}}{\cos \left(\pi \beta \right)}\, .
\end{align}
The function~(\ref{eq:solvpowerlaw}) is now completely
specified. Using the asymptotic behavior of the Bessel functions in
the limit $k\eta /\gamma \rightarrow 0$, one obtains the power
spectrum from Eq.~(\ref{eq:defP})
\begin{equation}
{\cal P}_{\zeta}=\frac{\gamma h(\beta )}{\pi \mpl^2\ell_0^2\epsilon _1}
\left(\frac{k}{\gamma}\right)^{2\beta +4}\, ,
\end{equation}
where the function $h(\beta)$ stands for
\begin{equation}
h(\beta)=\frac{\pi}{2^{2\beta +2}\cos ^2\left(\pi \beta \right)
\Gamma ^2\left(\beta +3/2\right)}\, .
\end{equation}
In the above expression, $\Gamma$ is the Euler function of first
kind. In the de Sitter case, $h(-2)=1$. Let us also notice that, in
the case of power-law inflation, the first slow-roll parameter
$\epsilon _1$ is actually constant and given by
\begin{equation}
\epsilon_1= \dfrac{2+\beta}{1+\beta}\,.
\end{equation}
As usual, for $\beta =-2$, one recovers $\epsilon _1=0$.

\par

We can now compare the previous expression with the one obtained in the
uniform approximation. The integral~(\ref{eq:fsup}) reads
\begin{equation}
  \dfrac{2}{3} f^{3/2}\left(k,\eta\right)
= \int _{\eta _*}^{\eta }{\rm d}\tau 
  \sqrt{\frac{(\beta +1/2)^2}{\tau
      ^2}-\frac{k^2}{\gamma ^2}}\, ,
\end{equation}
where, as already mentioned above, $\gamma $ is, in the present
context, a constant. This integral can be performed exactly. Then,
straightforward calculations lead to
\begin{equation}
{\cal P}_{\zeta}=\frac{\gamma j(\beta)}{\pi \mpl ^2\ell _0^2\epsilon _1}
\left(\frac{k}{\gamma }\right)^{2\beta +4}\, ,
\end{equation}
where the function $j(\beta)$ can be expressed as
\begin{equation}
j(\beta )=\frac{2{\rm e}^{2\beta +1}}{(2\beta +1)^{2\beta +2}}\, .
\end{equation}
The above power spectrum is the analogue of Eq.~(58) in
Ref.~\cite{Martin:2002vn}. In particular, the function $j(\beta )$ is
exactly the same as in the standard case. Therefore the error in the
amplitude of the power spectrum in the WKB/uniform approximation is
similar (see Fig.~1 of Ref.~\cite{Martin:2002vn}). In particular, and as
already mentioned, it is around $10\%$ for $\beta \sim -2$, \ie close to
scale invariance. For $\beta <-2$, when the spectrum is not necessarily
close to the Harrisson-Zeldovitch spectrum, the accuracy of the
WKB/uniform approximation becomes better. Let us also mention that
Ref.~\cite{Casadio:2004ru} has discussed how to improve the precision on
the amplitude and the method developed in that article may be applied
here. As can be noticed on the above equations, it turns out that the
spectral index is predicted exactly in the case of (DBI) power-law
inflation.

\subsection{Brane Inflation}
\label{sec:braneinf}

We now apply our formalism to the case of the brane inflationary models
discussed in Refs.~\cite{Chen:2004gc, Chen:2005ad,Bean:2006qz,
Chen:2006hs,Bean:2007eh}. In these scenarios, the warp function and
potential are given by the following expressions
\begin{equation}
\label{eq:pot-const}
T(\varphi)=\frac{\varphi^{4}}{\lambda}\, ,\quad
V(\varphi)=\dfrac{V_{0}}{1+\left(\dfrac{\mu}{\varphi}\right)^4}
+ \dfrac{\varepsilon}{2}\,m^{2}\varphi^{2}\, .
\end{equation}
Here, the factor $\varepsilon$ stands for $\varepsilon =\pm 1$ and the
positive sign corresponds to the so-called ``Ultra-Violet'' (UV)
models while the minus sign refers to the ``Infra-Red'' (IR)
scenarios~\cite{Bean:2007eh}. The Coulomb potential is due to the
attraction between a $\bar{D}3$-brane sitting at the bottom of the
throat and a mobile $D3$-brane, for a review see
Ref.~\cite{Lorenz:2007ze}. The quadratic correction in
Eq.~(\ref{eq:pot-const}) has a different status. It is a
phenomenological description of the brane moduli potential and its
shape is not established as neatly as for the Coulomb potential. The
above described model is characterized by four parameters, the mass
$m$, the scale $\mu $, the dimensionless 't~Hooft constant $\lambda $
and $V_0$ (whose dimension is $\mpl^{4}$). In the following it will
become clear that the evolution of the field is in fact controlled by
two dimensionless parameters $\alpha $ and $\beta$. The parameter
$\alpha $ is defined by
\begin{equation}
\label{eq:alpha}
\alpha \equiv \frac{12\pi \mpl ^2}{\lambda m^2}
=\frac{96\pi ^2}{\kappa \lambda m^2}\, , 
\end{equation}
while the other dimensionless parameter $\beta $ is given by
\begin{equation}
\beta \equiv \dfrac{V_0}{m^2\mpl^2}\, .
\end{equation}
Physically, $\beta$ measures the importance of the constant term
relative to the mass term in the potential.

\par

For those models, the vacuum expectation value (vev) of the inflaton
field possesses a geometrical interpretation, namely the distance
between the two branes living in the throat. As a consequence, it can
be written as $\varphi =\sqrt{T_3}r$, where
$T_3=1/\left[\left(2\pi\right)^3g_{_{\rm S}}\alpha '^2\right]$ is the
tension of the brane, $g_{_{\rm S}}$ being the string coupling,
$\alpha '$ the string scale and $r$ the radial coordinate along the
throat. Notice that we do not consider a possible angular motion of
the brane~\cite{Easson:2007dh,Easson:2007fz}. In order for the moving
brane to be inside the throat, one should impose $r<r_{\UV}$ where
$r_{\UV}$ is the coordinate at which the throat is connected to the
bulk. This quantity can be written as~\cite{Baumann:2006cd}
\begin{equation}
r_{\UV}^4=4\pi g_{_{\rm S}}\alpha '^2\frac{{\cal N}}{v}\, ,
\end{equation}
where ${\cal N}$ is a positive integer representing the total
Ramond--Ramond (RR) charge and $v$ represents the dimensionless ratio of
the five-dimensional volume forming the basis of the six-dimensional
conifold geometry to the volume of the unit five-sphere. Requiring the
volume of the throat to be smaller than the total volume of the
extra-dimensions amounts to
\begin{equation}
\varphi <\varphi_{\UV}<\frac{\mpl}{\sqrt{2\pi {\cal N}}}\, ,
\end{equation}
and inflation always occurs for sub-Planckian values of $\varphi$. On
the other hand, the bottom of the throat is located at $r_0$ and,
hence, one must have $\varphi>\varphi_0\equiv
\sqrt{T_3}r_0$. Moreover, for the model to be valid, the (physical)
distance between the brane must be larger than the string length and
one can show that this amounts to
\begin{equation}
\varphi >\varphi_{\rm strg}\equiv \varphi _0
{\rm e}^{\sqrt{\alpha'}{r_{\UV}}}\, .
\end{equation}

\par

Notice also that the parameters of the potential~(\ref{eq:pot-const})
can be calculated in terms of the stringy parameters. The tension
$T(\varphi)$, in the model under consideration, can also be written as
$T(\varphi)=T_3 (\varphi/\varphi _{\UV})^4$ which implies
that~\cite{Bean:2007eh}
\begin{equation}
\lambda =\frac{{\cal N}}{2\pi ^2v}\, .
\end{equation}
The constant term $V_0$ is given by $V_0=4\pi^2v\varphi_0^4/{\cal N}$ which
can also be expressed as
\begin{equation}
V_0=2h^4(r_0)T_3\, ,
\end{equation}
where $h(\varphi)\equiv \varphi /\varphi_{\UV}$ is the warp factor. On
the other hand, the constant $\mu$ can be expressed as $\mu^4=\varphi
_0^4/{\cal N}$.

\par

Finally, from the requirement that the volume of the throat is smaller
than the total volume of the extra-dimensions, one deduces
that~\cite{Bean:2007eh}
\begin{equation}
\label{eq:conditionba}
\sqrt{\frac{\beta
 }{\alpha}}<\frac{1}{\sqrt{24\pi^3}}\frac{h^2(r_0)}{{\cal N}}\ll 1\, .
\end{equation}

\par

We now study the different types of inflation that can occur in the
scenario under consideration. Each type corresponds to a different
choice of the parameters characterizing the model as discussed in
Ref.~\cite{Bean:2007eh} where the overall situation is summarized on two
phase diagrams.

\subsubsection{KKLMMT model}
\label{sec:kklmmt}

The quadratic term in Eq.~(\ref{eq:pot-const}) can be neglected in the
case where $V_0/m^2 \gg \mpl^2 >\varphi ^2$ or, in other words, for
$\beta \gg 1$. This, in turn, implies that $\alpha \gg 1$ because of
Eq.~(\ref{eq:conditionba}). This limit corresponds to the KKLMMT
model~\cite{Kachru:2003sx, Lorenz:2007ze}. The trajectory given by
Eq.~(\ref{eq:trajectory2}) reads
\begin{equation}
\label{eq:kklmmttrajectory}
  N(\varphi)=-\kappa\int^{\varphi}\dd
  \psi \sqrt{\frac{2}{3\kappa} + \frac{\mu^{2}}{16} \left(\frac{\psi}
      {\mu}\right)^{10}
    \left[1+\left(\frac{\mu}{\psi}\right)^{4}\right]}\, .
\end{equation}
The field rolls down the potential (\ie $\varphi $ is decreasing) while
inflation is under way and we always have $\varphi > \mu$. Therefore,
the previous expression can be approximated by
\begin{align}
  N(\varphi) & \simeq -\frac{\kappa\mu^{2}}{4}
  \bigg[\frac{1}{6}\left(\frac{\varphi} {\mu}\right)^{6} +\frac{1}{2}
  \left(\frac{\varphi}{\mu}\right)^{2}
  - \frac{1}{4} \left(\frac{\mu}{\varphi}\right)^{2} \nonumber\\
  & - \frac{4} {3\kappa\mu^{2}} \left(\frac{\mu}
    {\varphi}\right)^{4}\bigg]_{\varphi_{\uini}}^{\varphi}.
\end{align}
For $\varphi \gg \mu$, the first two terms in this expression clearly
dominate. In fact, this is precisely the trajectory found in
Ref.~\cite{Lorenz:2007ze} in the limit where the DBI dynamics can be
ignored. Therefore, the last two terms can be understood as the DBI
corrective terms induced by the non-standard kinetics. Clearly, these
corrections do not play an important role in the case of KKLMMT as
long as $\epsilon_1 \ll 1$. As shown in Ref.~\cite{Lorenz:2007ze}, the
DBI effects can only be significant for $\epsilon_1 > 1$ and we
therefore do not proceed further with the KKLMMT model\footnote{It was
  also shown in that reference that, depending on the values of the
  string parameters $\alphas$ and $\gs$ (\ie the string coupling), the
  end of inflation can occur either by violation of the slow-roll
  conditions or by instability. In the first case, the spectral index
  is given by $\nS\simeq 0.97$ while it is $\nS\simeq 1$ in the other
  situation.}.

\subsubsection{Chaotic Klebanov-Strassler models}
\label{sec:chaotic}

We now move on to the model where the quadratic correction dominates
(hence $\varepsilon =+1$) the potential. This corresponds to the
condition $\beta \ll 1$ and to the scenario considered in
Refs.~\cite{Silverstein:2003hf, Alishahiha:2004eh}, which we refer to as
chaotic Klebanov--Strassler (CKS) inflation. The potential and warp
function in this case read
\begin{equation}
\label{eq:chaoticKS}
V(\varphi)=\frac{1}{2}\,m^{2}\varphi^{2},\qquad 
T(\varphi)=\frac{\varphi^{4}}{\lambda}\, .
\end{equation}
The two remaining parameters are the mass $m$ and the dimensionless
constant $\lambda$. The integration of Eq.~(\ref{eq:trajectory2}) can
be performed explicitly and gives
\begin{eqnarray}
\label{eq:cks_srtraj}
\frac{\alpha ^{1/2}}{2\pi}N &=& -\sqrt{1+x^4}+\sqrt{1+x_{\rm ini}^4}
-2 \ln\dfrac{x}{x_{\rm ini}} \nonumber \\ & &
+\ln \left(\frac{1+\sqrt{1+x^4}}{1+\sqrt{1+x_{\rm ini}^4}}\right) ,
\end{eqnarray}
where we have defined 
\begin{equation}
x^4\equiv \alpha (\varphi/\mpl)^4.
\end{equation}
This is the implicit slow roll trajectory; in general, inverting this
expression to find $\varphi(N)$ is analytically impossible. However,
$\varphi$ decreases during inflation, therefore the initial field
value $\varphi_{\rm ini}$ is necessarily larger than all the field
values $\varphi$ attained during inflation. Assuming $\varphi \ll
\varphi_{\rm ini}$, an approximate inversion gives
\begin{eqnarray}
\label{eq:chaotictrajec}
  \varphi \simeq \varphi _{\rm ini}\exp \left( -\alpha^{1/2}
    \dfrac{N}{4\pi} \right) .
\end{eqnarray}
This expression is nothing but the solution $\varphi \rightarrow 1/t$
found in Ref.~\cite{Silverstein:2003hf} but expressed in terms of the
number of e-folds.

\par

We can also derive the behavior of $\gamma$, $\epsilon_{i}$ and
$\delta_{i}$. Using Eq.~(\ref{eq:gamma3}) together with
Eq.~(\ref{eq:Hslowroll}) in the limit where $\epsilon_{1}\ll1$, we find
the following expression for the Lorentz factor:
\begin{equation}
\label{eq:cks_gammasr}
\gamma(\varphi) \simeq \left(\frac{\mpl}{\varphi}\right)^2
\sqrt{\frac{\varphi^4}{\mpl^4}+\frac{1}{\alpha}}\,.
\end{equation}
Note that $\gamma\rightarrow\infty$ as $\varphi\rightarrow0$, \ie the
ultra-relativistic limit is attained at late times. Likewise, with the
help of Eq.~(\ref{eq:Hslowroll}), we find for $\epsilon_{1}$
\begin{equation}
\label{eq:cks_eps1sr}
  \epsilon_{1} \simeq \frac{1}{4\pi}\left(\frac{\varphi ^4}{\mpl^4}
    +\frac{1}{\alpha}
  \right)^{-1/2} .
\end{equation}
Note that $\epsilon_{1}$ approaches a constant as
$\varphi\rightarrow0$
\begin{equation}
  \lim_{\varphi \rightarrow 0} \epsilon_{1} =  \dfrac{\sqrt{\alpha}}{4\pi}\,.
\end{equation}
We recover that, in order to have slow-roll inflation for small values
of $\varphi$, the parameters of the model, $m$ and $\lambda $, must be
such that $\alpha \ll 1$.

\par

The above expression has also an important consequence with respect to
the end of inflation in this model. Indeed, the condition $\epsilon_1=1$
reduces to 
\begin{equation}
\frac{\varphi_{\rm end}}{\mpl}=\left(\frac{1}{16\pi ^2}-\frac{1}{\alpha}
\right)^{1/4}\, .
\end{equation}
Since we are in the limit $\alpha \ll 1$, this means that inflation
cannot stop by violation of the slow-roll conditions. It must be
stopped by another mechanism, typically by instability, comparable to
the case of standard hybrid inflation. The end of inflation $\varphi
_{\rm end}$ hence becomes a free, additional parameter of the model,
which is notably crucial for normalizing predictions to CMB
observations~\cite{Martin:2006rs}.

\par

\begin{figure}
\includegraphics[width=0.5\textwidth,clip=true]{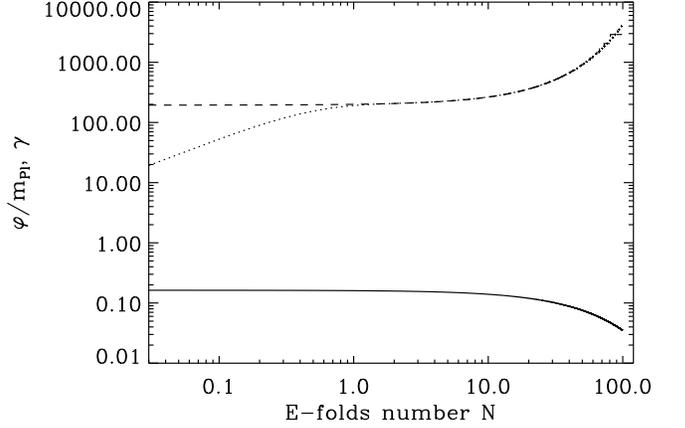}
\caption{Evolution of the scalar field $\varphi$ (solid line) and of
  the Lorentz factor $\gamma $ (dotted and dashed lines) as a function
  of the number of e-folds $N$, for $\alpha \simeq 0.04$ ($m=0.01\mpl
  $, $\lambda =10^7$). The initial value of the inflaton field is such
  that $\gamma _{\rm ini }= 3$. The solid line represents the
  trajectory of the inflaton field computed from the exact equation of
  motion. The DBI slow-roll trajectory of Eq.~(\ref{eq:cks_srtraj})
  cannot distinguished from the exact one on this plot. Concerning
  $\gamma$, except during a short transient regime at the beginning of
  inflation, the two curves also match.}
\label{fig:cks_field}
\end{figure}

\begin{figure}
\begin{center}
\includegraphics[width=0.5\textwidth,clip=true]{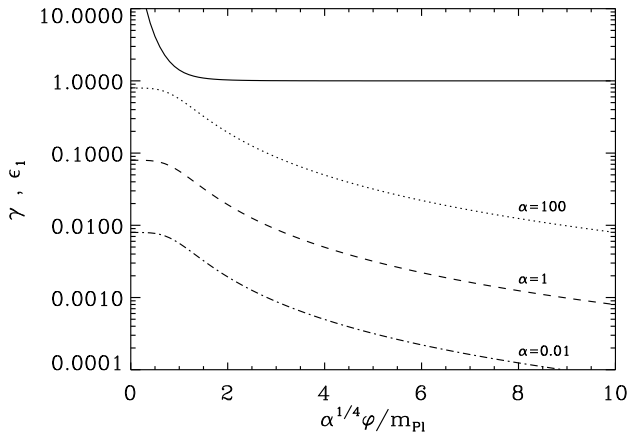}
\caption{Exact numerical evolution of the Lorentz factor $\gamma$
  (solid line) and $\epsilon_{1}$ as a function of the renormalized
  field value $\alpha^{1/4}\varphi/\mpl$ and for various values of the
  parameter $\alpha$. Notice that inflation proceeds from larger
  towards smaller field values. As can be seen from
  Eq.~(\ref{eq:cks_gammasr}), the $\alpha$-dependence of the Lorentz
  factor $\gamma$ is in the renormalized field. One can check that the
  asymptotic values of $\epsilon_1$ are compatible with the slow-roll
  predictions given in Eq.~(\ref{eq:cks_eps1sr}) (see also
  Fig.~\ref{fig:cks_srparam}).}
\label{fig:cks_srtrajec}
\end{center}
\end{figure}

\begin{figure}
\begin{center}
\includegraphics[width=0.5\textwidth,clip=true]{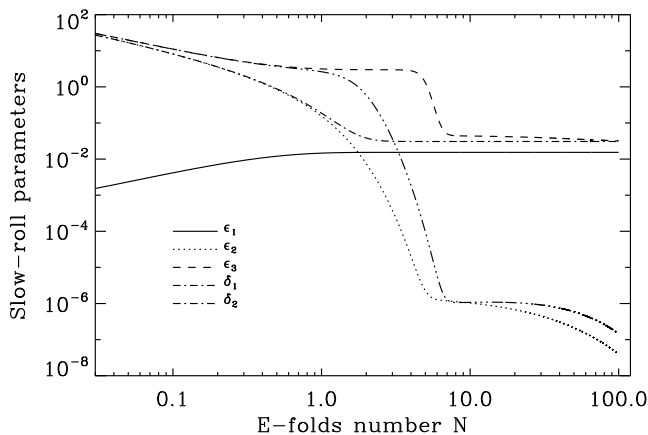}
\caption{Exact numerical evolution of the slow-roll parameters as a
  function of the number of e-folds $N$. The other parameters are the
  same as in Fig.~\ref{fig:cks_srtrajec}.}
 \label{fig:cks_srparam}
\end{center}
\end{figure}

The accuracy of the DBI slow-roll trajectory can be assessed by an exact
numerical integration of the equations of motion. The results are
plotted in Fig.~\ref{fig:cks_field} and confirm that, except during a few
e-folds after the beginning of inflation, the DBI slow-roll
approximation is accurate.
Similarly, the other Hubble flow functions can be approximated by
\begin{eqnarray}
\epsilon_{2}&\simeq&\frac{1}{2\pi}\left(\frac{\varphi}{\mpl}\right)^4
\left(\frac{\varphi ^4}{\mpl^4}+\frac{1}{\alpha}
\right)^{-3/2} ,\\
\epsilon_{3}&\simeq &-\frac{1}{\pi}\left(-\frac12\frac{\varphi ^4}{\mpl^4}
+\frac{1}{\alpha}\right)\left(\frac{\varphi ^4}{\mpl^4}+\frac{1}{\alpha}
\right)^{-3/2} ,
\end{eqnarray}
where Eq.~(\ref{eq:approxH}) has been used. As a result,
\begin{equation}
\label{eq:cksepslim}
  \lim_{\varphi \rightarrow 0} \epsilon_2 = 0, \qquad \lim_{\varphi
    \rightarrow 0 } \epsilon_3 =
  -\dfrac{\sqrt{\alpha}}{\pi} \sim -4\epsilon_{1}.
\end{equation}
Concerning the sound flow functions, one gets
\begin{eqnarray}
\delta_{1}&\simeq & \frac{1}{2\pi \alpha}\left(\frac{\varphi ^4}{\mpl^4}
+\frac{1}{\alpha}
\right)^{-3/2} ,\\
\delta_{2}&\simeq & \frac{3}{2\pi}\left(\frac{\varphi}{\mpl}\right)^4
\left(\frac{\varphi ^4}{\mpl^4}
+\frac{1}{\alpha}
\right)^{-3/2} .
\end{eqnarray}
Similarly, for small values of $\varphi$
\begin{equation}
\label{eq:cksdellim}
\lim_{\varphi \rightarrow 0}\delta_{1} = \dfrac{\sqrt{\alpha}}{2\pi}
\sim 2\epsilon _1, \qquad \lim_{\varphi\rightarrow 0 } \delta_2 = 0.
 \end{equation}
The evolution of the DBI slow-roll parameters is represented in
Fig.~\ref{fig:cks_srtrajec} and Fig.~\ref{fig:cks_srparam} and compared
to an exact numerical integration of the equations of motion. Again,
apart during the few e-folds of the transient regime, the DBI slow-roll
trajectory and parameters are in good agreement with the exact
integration.

As already noticed before and as clearly shown in
Figs.~\ref{fig:cks_field} and~\ref{fig:cks_srtrajec}, the Lorentz factor
$\gamma $ goes to infinity while $\varphi \rightarrow 0$. In this case,
one may question the use of the inflationary flow formalism (besides the
fact that it is hard to interpret since there is no natural measure in
parameter space) where the Hubble scale $H(\varphi)$ and $\gamma
(\varphi)$ are expanded in terms of the inflaton field at a specific
order~\cite{Peiris:2007gz}.

\par

Owing to the above analytical expressions, we can now easily consider
the observational predictions of this scenario. The multipole moments at
large angular scales are given by
\begin{equation}\label{eq:Cls}
C_{\ell}\simeq \frac{2H^2\gamma}{25\mpl^2\epsilon_1}
\frac{1}{\ell (\ell +1)}\, ,
\end{equation}
which, for $\ell=2$, gives the quadrupole moment
\begin{equation}
\label{eq:cobe}
\dfrac{Q}{\Tcmb} = \sqrt{\dfrac{5C_2}{4\pi}} \simeq 6\times 10^{-6}.
\end{equation}
The various quantities appearing in the above expressions should be
evaluated at ``sound horizon crossing''. Using the
trajectory~(\ref{eq:chaotictrajec})
\begin{equation}
\varphi _*=\varphi _{\rm end}\ue ^{\sqrt{\alpha} N_*/(4 \pi)},
\end{equation}
where $N_*$ is the number of e-folds between the end of inflation and
the sound horizon crossing. It is usually considered that
$40<N_*<60$. However, for the model under consideration, we have
$\gamma \propto 1/\varphi ^2$ and $H^2\propto \varphi ^2$ and,
therefore, these two terms cancel out of Eq.~(\ref{eq:Cls}). Since
$\epsilon_1\sim \sqrt{\alpha}/(4\pi)$, we have in fact the remarkable
property that the multipole moments do not depend on the number of
e-folds between sound horizon crossing and the end of inflation. This
property was first noticed in Ref.~\cite{Chen:2006hs}. Working out the
previous expressions, one arrives at
\begin{equation}
\dfrac{1}{\alpha}\left(\dfrac{m}{\mpl}\right)^2 = \dfrac{45}{4\pi}
\dfrac{Q^2}{\Tcmb^2} \simeq 1.3\times 10^{-10} .
\end{equation}
Since $\alpha \ll 1$, this means that the mass $m$ should be smaller
than in the standard chaotic scenario. The previous expression can be
re-expressed in terms of the parameter $\lambda$,
\begin{equation}
  \lambda \left(\dfrac{m}{\mpl}\right)^4 = 
125 \dfrac{Q^2}{\Tcmb^2} \simeq 0.0083.
\end{equation}

\par

Inserting Eqs.~(\ref{eq:cksepslim}) and (\ref{eq:cksdellim}) into the
expression of the scalar spectral index derived in Sec.~\ref{sec:index},
one obtains
\begin{equation}
\nS-1 =\order{\epsilon^{3},\delta^{3},
\epsilon^{i}\delta^{j}},\qquad i+j=3.
\end{equation}
The spectral index vanishes at first order in agreement with the
results of Refs.~\cite{Alishahiha:2004eh, Kinney:2007ag}. Here, we find
that, for this model, this also the case at second order. Notice
also that from Eq.~(\ref{eq:alpha}), the second and third-order
contribution in the running of the spectral index vanishes for
$\varphi\rightarrow0$ and $\alphaS$ reads
\begin{equation}
\alphaS = \order{\epsilon^{4},\delta^{4},
\epsilon^{i}\delta^{j}},\qquad i+j=4.
\end{equation}
Let us notice however that, although extremely small, the actual value
of $\alphaS$ obtained from the DBI slow-roll approximation is
significantly different than the value obtained from the numerical
integration, typically at almost $100\%$. Such a loss of accuracy is
related to the extremely flat power spectrum which has an almost
vanishing running; as for de Sitter, such a limit is in fact singular
for the scalar modes. In any case, such a running is by far too small
to be detectable in any of the present and planned cosmological
experiments.

\par

Then, one can compare the above predictions with the
WMAP5~\cite{Gold:2008kp, Hill:2008hx, Hinshaw:2008kr, Nolta:2008ih,
Dunkley:2008ie, Komatsu:2008hk} constraints on the CMB power spectra
established in Ref.~\cite{Lorenz:2008je}. Notice that, in order to
perform this comparison, it is mandatory to use a theoretical power
spectrum which takes into account the fact that the sound speed is a
time-dependent quantity. It would be inconsistent to do this comparison
using the constraints on the ordinary slow-roll parameters established,
for instance, in Refs.~\cite{Martin:2006rs,Kinney:2006qm}. From
Ref.~\cite{Lorenz:2008je}, we see that, in absence of a significant
contribution of gravitational waves, a red tilt is favored. Therefore,
the present model which is scale-invariant to a very high accuracy is
clearly disfavored by the current data.

\par

Let us remind that the model is also strongly disfavored by the WMAP
non-gaussianity bounds, as shown in
Ref.~\cite{Alishahiha:2004eh}. Indeed, at leading order for DBI
models, the parameter $\fNL$ is given by~\cite{Seery:2005wm,
  Chen:2006nt}
\begin{equation}
\fNL =  \frac{35}{108}\left(1-\gamma ^2\right)
\simeq -\frac{35}{108}\gamma ^2
\end{equation}
which, in the present context, gives
\begin{equation} 
\fNL\simeq -\frac{35}{108
 \alpha}\left(\frac{\mpl}{\varphi}\right)^4.
\end{equation}
The constraint $\varphi <\varphi _{\UV}$ implies~\cite{Bean:2007eh}
\begin{equation} 
  \fNL> -\frac{35\pi^2}{27 }
  \frac{{\cal N}^2}{\alpha} \gg 1 ,
\end{equation} 
because ${\cal N}>1 $ and $\alpha \ll 1$. The above inequality is in
contradiction with the observational range $-151\leq \fNL \leq 253$ (see
Ref.~\cite{Komatsu:2008hk}).

\subsubsection{Chaotic Klebanov Strassler models with constant term}
\label{sec:constantKS}

\begin{figure}
\begin{center}
\includegraphics[width=0.5\textwidth,clip=true]{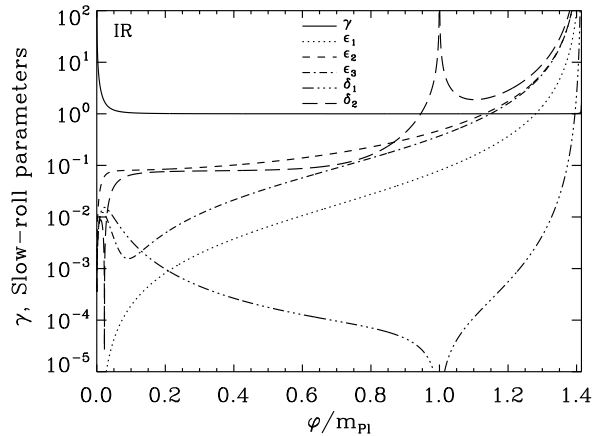}
\caption{Evolution of the Lorentz factor $\gamma $ and of the Hubble
  and sound flow functions in the IR case of the CKS models with
  constant term. The parameters are $\alpha =500$ and $\beta
  =1$. Inflation proceeds from left to right and the ``divergence''
  occurs at $\varphi/\mpl \rightarrow \sqrt{2\beta }$.}  \label{fig:ir}
\end{center}
\end{figure}

We now consider the case where the Coulomb term is negligible and the
potential can be written as
\begin{equation}
\label{eq:potwocoulomb}
V(\varphi)=V_{0}+ \dfrac{\varepsilon}{2}\,m^{2}\varphi^{2} ,
\end{equation}
while the warp function remains unchanged and given by its expression in
Eq.~(\ref{eq:pot-const}). These models have been discussed in
Refs.~\cite{Chen:2004gc,Chen:2005ad,Bean:2006qz,
Chen:2006hs,Bean:2007eh}.

\par

\begin{figure}
\begin{center}
  \includegraphics[width=0.5\textwidth,clip=true]{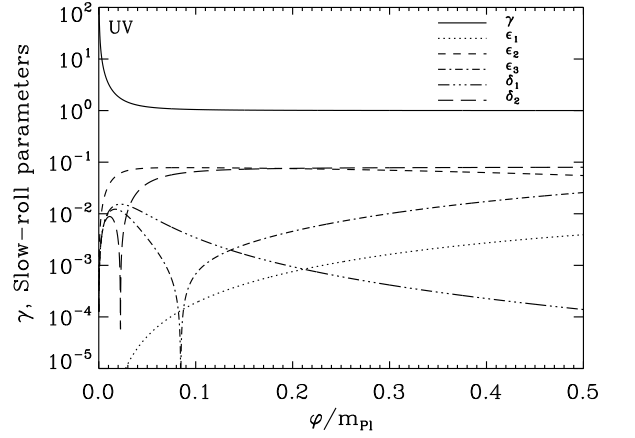} \caption{Same
    as Fig.~\ref{fig:ir} but in the UV case. The inflaton field
    decreases while inflation is under way.} \label{fig:uv}
\end{center}
\end{figure}

Using Eqs.~(\ref{eq:phidotdbi}) and~(\ref{eq:approxH}) leads to
$\dot{\varphi}=-\varepsilon m^2\varphi /(3\gamma H)$. This means that,
in the UV ($\varepsilon=1$) case, the vev of the inflaton field is
decreasing as inflation proceeds while it is increasing in the IR case
($\varepsilon=-1$). In order to understand in which regimes DBI
inflation can occur, it is convenient to express the Lorentz factor in
terms of the vacuum expectation value of the inflaton field as
\begin{equation}
\gamma\left(\varphi\right)
= \left(\frac{\mpl}{\varphi}\right)^2
\left[\left(\frac{\varphi}{\mpl}\right)^4+\frac{1}{\alpha }
\frac{\left(\varphi/\mpl\right)^2}{2\beta 
+\varepsilon \left(\varphi/\mpl\right)^2}\right]^{1/2}.
\end{equation}
Of course, when $\varepsilon=+1$ and $\beta =0$, one checks that
Eq.~(\ref{eq:cks_gammasr}) is recovered. The DBI regime, $\gamma
\rightarrow +\infty$, is obtained when $\varphi \rightarrow 0$. In this
case, $\gamma$ can be approximately written as
\begin{equation}
\gamma {\simeq} \frac{1}{\sqrt{2\alpha \beta}}\frac{\mpl}{\varphi}\, .
\end{equation}
Let us notice that this last equation is valid for $\varepsilon=\pm
1$. In addition, in the IR case, the Lorentz factor also blows up when
$\varphi \rightarrow \mpl \sqrt{2\beta}$ as
\begin{eqnarray}
\gamma_{\IR} {\simeq} \frac{1}{\sqrt{2\alpha\beta}}
\left[2\beta -\left(\frac{\varphi}{\mpl}\right)^2\right]^{-1/2}\, .
\end{eqnarray}
The evolution of $\gamma (\varphi)$ is displayed in
Figs.~\ref{fig:ir} and~\ref{fig:uv}. Notice that we must always have
$\varphi <\mpl \sqrt{2\beta}$ in order to guarantee the positivity of
the potential. Moreover, since in the limit $\varphi \to \mpl
\sqrt{2\beta}$, $V(\varphi) \rightarrow 0$, the corrections to the
potential may become non-negligible. In particular, the Coulomb part of
the potential that we have neglected in this subsection will become
important again. Notice that this situation is rather similar to the
case of small field models in standard inflation where the potential is
given by $V\propto 1-\phi^2$ and where, at the end of the slow-roll
phase, it is necessary to add higher order terms of the form $\phi ^p$
in order for the potential to have a minimum which ensures that the
reheating can proceed. We conclude that the limit $\varphi \rightarrow
\mpl \sqrt{2\beta}$, although it does imply $\gamma \rightarrow
+\infty$, appears to be more difficult to realize from the physical
point of view.

\par

For all $\varepsilon$, the first slow-roll parameter can be written as
\begin{equation}
\label{eq:eps1dbiconst}
\epsilon_1\left(\varphi\right) = \frac{1}{4\pi \gamma\left(\varphi\right)}
\frac{\left(\varphi/\mpl\right)^2}
{\left[2\beta +\varepsilon \left(\varphi/\mpl\right)^2\right]^2}\,,
\end{equation}
which, in the limit $\varphi \rightarrow 0$, becomes
\begin{equation}
\label{eq:eps1UVzero}
\epsilon _1\simeq \frac{1}{16\pi}\sqrt{\frac{2\alpha}{\beta ^3}}
\left(\frac{\varphi}{\mpl}\right)^3,
\end{equation}
and goes to zero as the inflaton vev vanishes. This means that, in the
UV case, where the field decreases from some initial value towards
zero, inflation cannot stop by violation of the slow-roll condition.

\par

On the contrary, in the IR case, the field starts from a value close to
zero and increases. As already mentioned, the vev of the field is
bounded by $\sqrt{2\beta}$. In this limit, the first slow-roll parameter
blows up as
\begin{equation}
\epsilon_1 \simeq \frac{\sqrt{8\alpha \beta ^3}}{4\pi}
\left[2\beta -\left(\frac{\varphi}{\mpl}\right)^2\right]^{-3/2},
\end{equation}
and contrary to the UV case, inflation will stop by violation of the
slow-roll conditions. The evolution of $\epsilon_1$ is displayed in
Fig.~\ref{fig:ir} (IR case) and in Fig.~\ref{fig:uv} (UV case).

\par

It is also interesting to establish the expression of the other
slow-roll parameters. Lengthy but straightforward calculations lead to
the following expressions
\begin{eqnarray}
\epsilon_2 &=& 4\epsilon_1-\delta _1 -\frac{\varepsilon \alpha}{2\pi }
\frac{\varphi^2}{\mpl^2}\frac{\gamma ^2-1}{\gamma}\, ,\\
\epsilon _3 &= & 4 \epsilon_1-\frac{\delta _1\delta_2}{\epsilon_2}
+\frac{\alpha ^2}{4\pi^2\epsilon_2}\frac{\varphi ^4}{\mpl^4}
\frac{\left(\gamma ^2-1\right)^2}
{\gamma ^2}\nonumber \\ &  - &\frac{\varepsilon \alpha}{2\pi}
\frac{\delta _1}{\epsilon_2}\frac{\varphi^2}{\mpl^2}\frac{\gamma
^2+1}{\gamma }\,, \\
\delta _1 &=& \frac{\varepsilon \epsilon_1}{\alpha \gamma ^2}
\frac{\mpl^4}{\varphi^4}\left[1+\varepsilon \alpha 
\frac{\varphi ^4}{\mpl^4}\left(\gamma ^2-1\right)\right] , \\
\delta_2 &=& \epsilon_2-2\delta _1+\frac{\varepsilon \alpha }{\pi \gamma
 }\left(\gamma ^2-1\right)\frac{\varphi ^2}{\mpl^2}
\nonumber \\ & + &
\varepsilon \alpha \gamma ^2 \frac{\varphi ^4}{\mpl^4}
\left[1+\varepsilon \alpha \left(\gamma ^2-1\right)
\frac{\varphi^4}{\mpl^4}\right]^{-1}
\nonumber \\ &  \times &
\left[2\delta _1-\frac
{\varepsilon \alpha}{\pi \gamma}\frac{\left(\gamma ^2-1\right)^2}
{\gamma ^2}\frac{\varphi ^2}{\mpl^2}\right] .
\end{eqnarray}
These functions are represented in Fig.~\ref{fig:ir} for the IR case
($\varepsilon=-1$) and in Fig.~\ref{fig:uv} in the UV case
($\varepsilon=+1$).

With a constant term included, Eq.~(\ref{eq:trajectory2}) describing
the field trajectory becomes significantly more complicated. Expressed
in terms of the parameters $\alpha$ and $\beta$, the DBI trajectory
reads
\begin{align}
  N(\varphi) =&- 2\pi\varepsilon\beta \int_{u_\uini}^{u} \ud u
\left(u-\varepsilon\right)^{-3/2}\nonumber\\
&\times\left[u^3+\varepsilon u^2
+\left(\frac{\varepsilon}{\alpha\beta^{2}}-1\right)u
+\frac{1}{\alpha\beta^{2}}-\varepsilon\right]^{1/2},\label{eq:traj-ell2}
\end{align}
where we have defined the new dimensionless variable $u$ by
\begin{equation}
\label{eq:defu}
u\equiv \frac{1}{\beta}\left(\frac{\varphi}{\mpl}\right)^2+\varepsilon .
\end{equation}
The expression (\ref{eq:traj-ell2}) is an elliptic integral and can be
expressed in terms of the canonical elliptic integrals of the first and
second kind, respectively denoted as in Ref.~\cite{Abramovitz:1970aa} by
$\ellF{\theta}{p}$ and $\ellE{\theta}{p}$. We will further use the
definition
\begin{equation}
\label{eq:parameters}
n \equiv \dfrac{\alpha\beta^{2}}{\alpha\beta^{2}-\varepsilon}\,.
\end{equation}
To ensure that all expressions are well-defined in the following, we now
consider separately the case of IR and UV models.

\par

\begin{figure}
\includegraphics[width=0.5\textwidth,clip=true]{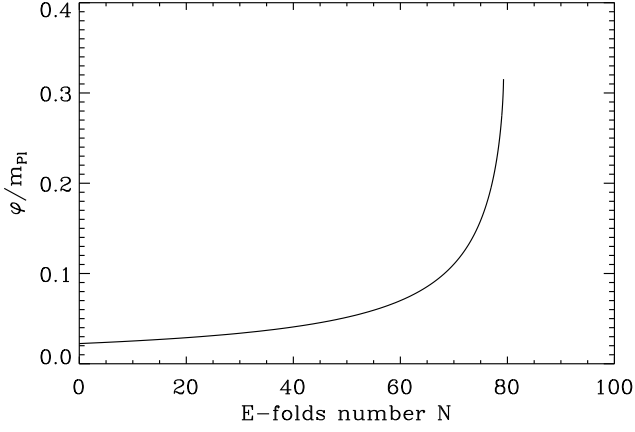}
\caption{Typical evolution of the scalar field $\varphi$ according to
Eq.~(\ref{eq:trajectoryIR}) in the IR case. The parameters are $\alpha
=4$ and $\beta =0.05$ such that the condition $\beta /\alpha \ll 1$ is
satisfied. The initial condition is $\varphi_{_{\rm ini}}/\mpl\simeq
0.022$ and one can check that the field value always remains
sub-Planckian. With the parameters considered here, one has $n=0.1$, \ie
$0<n<1$ as required for the IR case.}  \label{fig:trajecir}
\end{figure}

For the IR case, $\varepsilon=-1$ and it follows from
Eq.~(\ref{eq:defu}) that the variable $u$ maps the range $\varphi/\mpl
\in\left[0, \sqrt{2\beta}\right]$ to $u\in[-1,1]$. From
Eq.~(\ref{eq:parameters}), we see that $0<n<1$ and both $u$ and $n$
appearing in Eq.~(\ref{eq:traj-ell2}) are in the canonical domain of
definition of the elliptic integrals $\ellE{\theta}{p}$ and
$\ellF{\theta}{p}$ (see Ref.~\cite{Abramovitz:1970aa}). One gets
\begin{widetext}
\begin{equation}
\label{eq:trajectoryIR}
\begin{aligned}
  N_{\IR}(\varphi)&= 2\pi\beta \left[ \frac{2}{\sqrt{n}}\ellF{\arcsin
      u}{n} -\frac{2}{\sqrt{n}} \ellF{\arcsin u_\uini}{n}
    -\frac{3}{\sqrt{n}} \ellE{\arcsin u}{n} +\frac{3}{\sqrt{n}}
    \ellE{\arcsin u_\uini}{n}
    \phantom{\sqrt{\frac{1/n-u^2_\uini}{1-u^2_\uini}}} \right.
  \\
  &
  \left. -2\ln\left(\frac{\sqrt{1-u_\uini^{2}}+\sqrt{1/n-u_\uini^{2}}}
      {\sqrt{1-u^{2}}+\sqrt{1/n-u^{2}}}\right)
    -2(1-u)\sqrt{\frac{1/n-u^{2}}{1-u^2}}
    +2(1-u_\uini)\sqrt{\frac{1/n-u^2_\uini}{1-u^2_\uini}} \right] .
\end{aligned}
\end{equation}
\end{widetext}
A typical IR trajectory is represented in Fig.~\ref{fig:trajecir}. 
This expression, although exact, is quite involved. Therefore, it is
interesting to consider the dominant behavior for $u\rightarrow-1$, or
$\varphi \rightarrow 0$, \ie for very early times. In this
situation, it turns out that the terms containing the square roots are
the dominant ones because they contain a pole. Then, neglecting the
other terms, one can express the variable $u$ and, hence, the inflaton
field vev $\varphi$ in terms of the number of e-folds. This leads to
\begin{equation}
\label{eq:trajecapproxir}
\varphi_{\IR}(N) \simeq \varphi _{\rm ini}
\left[1-\frac{\varphi_{\rm ini}}{\mpl}\sqrt
{\frac{\alpha}{32\pi^2\beta}}N\right]^{-1} .
\end{equation}
One checks that when the number of e-folds increases, the vev of the
inflaton field increases. This clearly reproduces the behavior seen in
 Fig.~\ref{fig:trajecir}. Notice that, in order to establish the
previous formula, we have also assumed that $\varphi_{\rm ini}/\mpl\ll
1$ since $\varphi /\mpl\ll 1$ and $\varphi _{\rm
  ini}<\varphi$. Moreover, the approximation is valid only when the
second term in the squared bracket of Eq.~(\ref{eq:trajecapproxir}) is
small in comparison to one such that the vev remains positive. When
$N$ is too large, the brane is far from the bottom of the throat and
this approximation breaks down. Finally, in the previous section, we
have established the constraint $\beta /\alpha \ll 1 $. We see that,
in Eq.~(\ref{eq:trajecapproxir}), the inverse of this factor appears
and in order to be consistent with the condition $\varphi/\mpl \ll 1$,
one must also have $\varphi_\uini/\mpl\ll \sqrt{\beta/\alpha}$.

\par

Let us now turn to the UV case. In this situation, the variable $u$
covers the range $u\in[1,+\infty[$ where the lower bound corresponds
to $\varphi \rightarrow 0$. This is out of the canonical domain of
definition for the elliptic integrals and to carry out the integration
of Eq.~(\ref{eq:traj-ell2}), one can introduce the variable
$1/u$. Moreover, the parameter $n$ has now two disjoint ranges: for
$0<\alpha\beta^{2}<1$, $n\in ]-\infty,0[$, whereas for
$\alpha\beta^{2}>1$  we get  $n\in]1,+\infty[$. Some care is
therefore required on the formulations of the elliptic integrals.

\par

We first focus on the case where $\alpha \beta^2>1$. After appropriate
redefinitions~\cite{Abramovitz:1970aa}, the final trajectory reads
\begin{widetext}
\begin{equation}
\label{eq:trajUVpos}
\begin{aligned}
  N_{\UV}^{(n>1)}(\varphi) & = 2\pi\beta \left[ \frac{3n-1}{n}
    \ellF{\arcsin \frac{1}{u}}{\frac{1}{n}} - \frac{3n-1}{n}
    \ellF{\arcsin \frac{1}{u_{\rm ini}}}{\frac{1}{n}} -3
    \ellE{\arcsin\frac{1}{u}}{\frac{1}{n}} +3 \ellE{\arcsin
      \frac{1}{u_{\rm ini}}}{\frac{1}{n}}
    \phantom{\sqrt{\frac{u^{2}_{\rm ini}-1/n}{u^{2}_{\rm ini}-1}}}
  \right.
  \\
  & \left. +2 \ln
    \left(\frac{\sqrt{u_{\uini}^{2}-1/n}+\sqrt{u_{\uini}^{2}-1}}
      {\sqrt{u^{2}-1/n}+\sqrt{u^{2}-1}}\right)
    -\frac{(u-1)^{2}-4}{u}\sqrt{\frac{u^{2}-1/n}{u^{2}-1}}
    +\frac{(u_{\rm ini}-1)^{2}-4}{u_{\rm ini}} \sqrt{\frac{u^{2}_{\rm
          ini}-1/n}{u^{2}_{\rm ini}-1}} \right] .
\end{aligned}
\end{equation}
\end{widetext}
At the end of inflation, \ie when $\varphi\rightarrow0$ (or
$u\rightarrow 1$), the next term to last in Eq.~(\ref{eq:trajUVpos})
dominates, and we can invert the resulting expression to obtain
\begin{equation}
\label{eq:trajecapproxuv}
\frac{\varphi_{\UV} (N)}{\mpl}\simeq \sqrt{\frac{32\pi ^2\beta}{\alpha}}
\frac{1}{N}\, .
\end{equation}
Let us notice that, this time, the factor $\sqrt{\beta /\alpha}\ll 1$
directly appears in the above expression and, therefore, guarantees the
consistency of the formula expressing the quantity $\varphi/\mpl\ll 1$.

\par

\begin{figure}
\includegraphics[width=0.5\textwidth,clip=true]{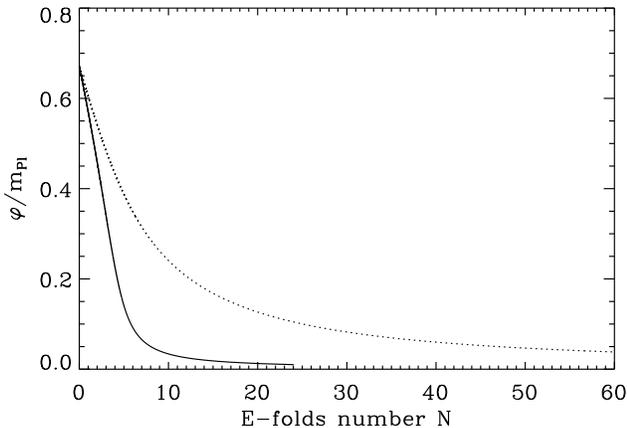}
\caption{Typical evolution of the scalar field $\varphi$ according to
Eq.~(\ref{eq:trajUVpos}) and Eq.~(\ref{eq:trajUVneg}) in the UV
cases. The solid line refers the UV case with $n>1$ since the parameters
chosen are $\alpha =404$ and $\beta=0.05$ which implies $n=101$. On the
contrary, the dotted line represents a trajectory in the UV case with
$n<0$. The corresponding parameters are $\alpha =4$, $\beta =0.05$ which
gives $n=-0.01$. In both cases, the initial condition is $\varphi_{_{\rm
ini}}/\mpl\simeq 0.67$.}  \label{fig:trajecuv}
\end{figure}

Let us now consider the case  $0<\alpha \beta^2<1$.   The
elliptic integrals $\ellF{\theta}{p},\,\ellE{\theta}{p}$ that arise in
the calculation of Eq.~(\ref{eq:traj-ell2}) have  now to be
redefined for complex arguments $\theta$ and/or negative parameters
$p$, respectively. After some calculations~\cite{Abramovitz:1970aa},
the trajectory reads
\begin{widetext}
\begin{equation}
\label{eq:trajUVneg}
\begin{aligned}
\frac{N_{\UV}^{(n<0)} (\varphi)}{2\pi \beta} & = 
\sqrt{1+\frac{1}{\abs{n}}}
\Biggl[
\ellF{\arcsin\sqrt{\frac{1+1/\abs{n}}
{u^{2}+1/\abs{n}}}}{\frac{1}{\abs{n}+1}}
-\ellF{\arcsin \sqrt{\frac{1+1/\abs{n}}{u^{2}_{\rm ini} 
+ 1/\abs{n}}}}{\frac{1}{\abs{n}+1}} \\
&- \ellE{\arcsin \sqrt{\frac{1+1/\abs{n}}
{u^{2}+1/\abs{n}}}} {\frac{1}{\abs{n}+1}} 
+ \ellE{\arcsin \sqrt{\frac{1+1/\abs{n}}{u^{2}_{\rm ini}+1/\abs{n}}}}
{\frac{1}{\abs{n}+1}}\Biggr] \\
& +2\sqrt{\frac{\abs{n}}{\abs{n}+1}}\Biggl[
-\ellF{\arcsin\frac{\sqrt{u^{2}-1}}{u}} {\frac{1}{\abs{n}+1}}
+\ellF{\arcsin \frac{\sqrt{u^{2}_{\rm ini}-1}}{u_{\rm ini}}} 
{\frac{1}{\abs{n}+1}} \\ 
& + \ellE{\arcsin \frac{\sqrt{u^{2}-1}}{u}} {\frac{1}{\abs{n}+1}}
- \ellE{\arcsin \frac{\sqrt{u^{2}_{\rm ini}-1}}{u_{\rm ini}}}
{\frac{1}{\abs{n}+1}} \Biggr] \\ 
&- 2 \ln \left(\frac{\sqrt{u^{2}-1}+\sqrt{u^{2}
+1/\abs{n}}}{\sqrt{u_{\uini}^{2}-1}
+ \sqrt{u_{\uini}^{2}+1/\abs{n}}} \right)
- \dfrac{u^{2}\left(u-1\right)^{2} - 4u^{2} - 2\left(1+u\right)/\abs{n}}
{u\sqrt{\left(u^{2}-1\right)\left(u^{2}+1/\abs{n}\right)}} \\
 & + \dfrac{u^{2}_{\rm ini}\left(u_{\rm ini}-1\right)^{2} 
-4u^{2}_{\rm ini}-2\left(1+u_{\rm ini}\right)/\abs{n}}
{u_{\rm ini}\sqrt{\left(u^{2}_{\rm ini}-1\right)\left(u^{2}_{\rm ini}
+1/\abs{n}\right)}}\, .
\end{aligned}
\end{equation}
\end{widetext}
In the limit $u\rightarrow1$ or $\varphi \rightarrow 0$, the next to
last term in Eq.~(\ref{eq:trajUVneg}) dominates and we can approximately
invert the above expression. The resulting trajectory ends up being
given by the same expression as in
Eq.~(\ref{eq:trajecapproxuv}). Finally, two typical UV trajectories are
represented in Fig.~\ref{fig:trajecuv}.

Using these results, let us now turn to the observational predictions
of these models. The Cosmic Background Explorer (COBE) normalization,
implemented with the help of Eq.~(\ref{eq:cobe}), leads to the
following relation, valid for the UV and IR cases
\begin{equation}
\left(\frac{\mpl}{\varphi_*}\right)^4 = \dfrac{45}{16 \pi}
\dfrac{Q^2}{\Tcmb^2} \left(\frac{\mpl}{m}\right)^2 \frac{\alpha }{\beta ^2}
\, .
\end{equation}
This formula is only consistent when $\varphi _*/\mpl $ is a small
quantity. In the limit $\varphi \rightarrow 0$, one has from
Eq.~(\ref{eq:eps1UVzero}) $\epsilon _1\propto \varphi ^3$ and
\begin{eqnarray}
  \delta _1 \underset{\varphi \to 0}{\sim} \frac{\varepsilon}{8\pi}
  \sqrt{\frac{2\alpha}{\beta}} \frac{\varphi}{\mpl} \, ,\qquad
  \epsilon _2 \underset{\varphi \to 0}{\sim}  -3\delta _1.
\end{eqnarray}
Therefore, $\epsilon _2$ and $\delta _1$ give the dominant
contribution to the spectral index and one has
\begin{equation}
\nS - 1 \simeq 4\delta _1.
\end{equation}
Using the COBE normalization given above leads to
\begin{equation}
\label{eq:ckscte_ns}
\nS-1 \simeq \varepsilon \left(\dfrac{16 \Tcmb^2}{15 \pi^2 Q^2} \right)^{1/4}
\lambda^{-1/4} \simeq 234.1 \varepsilon \times \lambda ^{-1/4}.
\end{equation}
The spectral index only depends on the dimensionless 't~Hooft coupling
constant $\lambda$ which is quite remarkable given the complexity of
the equations and the number of free parameters of the model. In the
UV case, the tilt is positive while it is negative in the IR case. In
order to compatible with the CMB data, one sees that one must have
$\lambda \gta 3\times 10^{13}$.

\par

The calculation of the running   proceeds along the same
lines. In the limit $\varphi \to 0$, one has
\begin{eqnarray}
  \delta _2 \underset{\varphi \to 0}{\sim} -\delta_1 ,\qquad
  \epsilon _3 \underset{\varphi \to 0}{\sim}  \frac13
\left(1-4\varepsilon\right)\delta _1.
\end{eqnarray}
Then, using Eq.~(\ref{eq:alphaSsecond}), one obtains the following
expression
\begin{equation}
\label{eq:ckscte_alpha}
\alphaS\simeq -4\varepsilon \delta _1^2=-\frac{\varepsilon}{4}
\left(\nS-1\right)^2.
\end{equation}
The IR models ($\varepsilon=-1$) have  therefore a red spectral
index and a positive running while, on the contrary, the UV models
($\varepsilon=1$) have a blue spectral index and a negative
running.
\begin{figure}
\begin{center}
\includegraphics[width=0.455\textwidth]{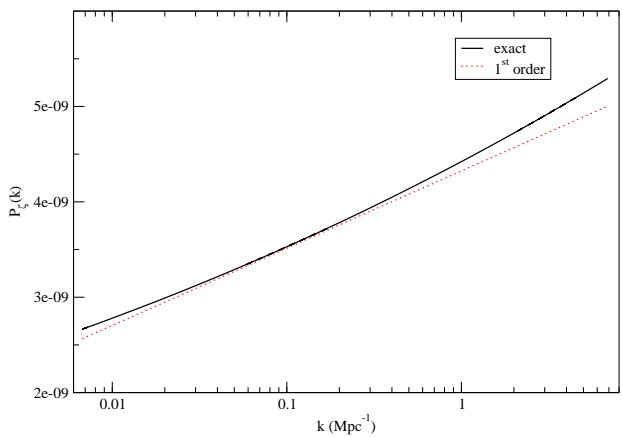}
\caption{Exact (solid) and first order (dotted) scalar power spectrum
  for the CKS plus constant UV models. The parameters are $\alpha=38$,
  $\beta=3.7$, $\lambda=1.6\times 10^{13}$ and $m^2=5\times
  10^{-4}\mpl^2$. From the exact spectrum, one gets at the pivot $\nS-1
  \simeq 0.11$ and $\alphaS \simeq -0.0023$ in agreement with
  Eqs.~(\ref{eq:ckscte_ns}) and (\ref{eq:ckscte_alpha}).}
\label{fig:cksctePk}
\end{center}
\end{figure}
 In Fig.~\ref{fig:cksctePk}, we have plotted the scalar power
spectrum for the UV model stemming from an exact numerical integration
of Eq.~(\ref{eq:eomv}). Both the spectral index and the running match
at a few percent with Eqs.~(\ref{eq:ckscte_ns}) and
(\ref{eq:ckscte_alpha}).

\par

One can also estimate the tensor to scalar contribution. For
this purpose, one can compute the parameter $r=16\epsilon _1/\gamma$,
see Eq.~(\ref{eq:tenstoscal}). Using again the COBE normalization,
straightforward manipulations lead to 
\begin{eqnarray}
r \simeq \dfrac{128 \pi \Tcmb^2}{15 Q^2} \dfrac{\beta}{\alpha} \lambda^{-1},
\end{eqnarray}
which is very small since $\beta/\alpha \ll 1$ and $\lambda^{-1} \propto
(\nS - 1)^4$. Finally, the level of non-gaussianity can be estimated to
\begin{equation}
  \fNL \simeq -\dfrac{35}{864} \sqrt{\dfrac{45}{\pi}} \dfrac{Q}{\Tcmb}
 \dfrac{\mpl/m}{\alpha^{1/2} \beta^{2}} \simeq  -9.2 \times
  10^{-7}
  \frac{\mpl/m}{\alpha ^{1/2}\beta ^2}\, .
\end{equation}
 which is negative.

\section{Conclusion}
\label{sec:conclusion}

In this section, we briefly recap and discuss the results obtained
before. The main goal of our article was to present a generic
formalism, as similar as possible to the standard approach, to
investigate the features and the observational predictions of
k-inflation. In particular, we have offered a new formula (but valid
for DBI inflation only) for computing the classical trajectory for any
warp function $T(\varphi)$ and potential $V(\varphi)$. Then, using the
uniform approximation, we have computed the scalar and tensor power
spectra, see Eqs.~(\ref{eq:scalarspectrum})
and~(\ref{eq:tensorspectrum}). These formulas are based on a double
expansion. There is first an expansion around a pivot scale and, then,
each coefficient of this expansion is in turn evaluated by means of
another expansion, this time in terms of the Hubble and sound flow
parameters. Finally, we have illustrated our results  on various
example models encountered in the literature.  In particular, we
have derived new results concerning the CKS models with constant term
and obtained simple expressions for the cosmological observables.

\par

Having generalized the standard formalism to k-inflation, the next
step is of course to compare the predictions to the observations. This
is done in Ref.~\cite{Lorenz:2008je} in the case of the CMB WMAP5
data.

\begin{acknowledgments}
  LL acknowledges support through a DAAD PhD scholarship. This work is
  partially supported by the Belgian Federal Office for Science,
  Technical and Cultural Affairs, under the Inter-university
  Attraction Pole grant P6/11.
\end{acknowledgments}

\begin{appendix}

\section{Potential and warp slow-roll parameters}
\label{app:srparam}

In a broader sense, one can also try to establish another hierarchy
of expansion parameters to the DBI case, namely in terms of the
potential $V(\varphi)$ and the geometry parameter $T(\varphi)$. If we
define the two parameters $\epsilon_{_{\rm V}}$ and $\epsilon_{_{\rm
T}}$ according to
\begin{equation}
\epsilon_{_{\rm V}}\equiv \frac{1}{2\kappa}\left(\frac{V'}{V}\right)^{2},
\qquad\epsilon_{_{\rm T}}\equiv \frac{1}{2\kappa}\left(\frac{T'}{T}\right)^{2},
\end{equation}
where, in the present context, a prime means a derivative with respect
to $\varphi$. We can establish the link with the
$\left(\epsilon_{i},\,\delta_{i}\right)$ parameters. Straightforward
manipulations lead to the following expressions
\begin{eqnarray}
\epsilon_{_{\rm T}} &=&\frac{\gamma}{\gamma^{2}-1}\frac{\delta_{1}^{2}}
{\epsilon_{1}}, \\
\epsilon_{_{\rm V}} &=&\frac{\epsilon_{1}\gamma}{4}\left[\frac{2(\gamma+1)
+\gamma(\delta_{1}-\epsilon_{2}-2\epsilon_{1})}
{1+\gamma(1-\epsilon_{1})}\right]^{2}.
\end{eqnarray}
As can be expected, there is no one-to-one correspondence between the
two sets of parameters (\ie, for instance, the expression of
$\epsilon_{_{\rm T}}$ not only involves $\epsilon_1 $ but also $\delta
_1$).

\section{Relations with other slow-roll hierarchies}
\label{app:relsrparam}

In this Appendix, we briefly state the relations between our hierarchy
of $\left(\epsilon_{i},\delta_{i}\right)$ parameters and the various
parameter sets that have been used elsewhere in the literature. The
parameters $\left(\epsilon_D,\eta_D,\kappa_D,\xi_D,\rho_D\right)$
defined in Ref.~\cite{Shandera:2006ax} can be expressed in terms of our
parameters as follows:
\begin{eqnarray}
  \epsilon_{D}&=&\epsilon_{1},\\
  \eta_{D}&=&\epsilon_{1}-\frac{1}{2}(\epsilon_{2}+\delta_{1}),\\
  \kappa_{D}&=&-\delta_{1},\\
  \xi_{D}&=&\frac{\epsilon_{2}\epsilon_{3}}{2} +\frac{\delta_{1}
    \epsilon_{2}} {2}-\frac{3}{2}\,\epsilon_{1}\epsilon_{2}\nonumber\\
  &&+\epsilon_{1}\left(\epsilon_{1}-\frac{3}{2}\, \delta_{1}\right)
  + \frac{\delta_{1}}{2}(\delta_{1}+\delta_{2}), \\
  \rho_{D}&=&\frac{\delta_{1}}{2\epsilon_{1}}(2\delta_{2}+3\delta_{1}
  - \epsilon_{2}).
\end{eqnarray}
In Ref.~\cite{Bean:2007hc}, the first three of these parameters were
used, changing the notation to $\epsilon_D\rightarrow\epsilon$ and so
on. Then, it is interesting to compare the expression~(2.30) of
Ref.~\cite{Shandera:2006ax} to our Eq.~(\ref{eq:ns}). Written in terms
of the Hubble and sound horizon parameters, this equation reads
\begin{eqnarray}
\nS-1 & = & -2\epsoneP-\epstwoP+\deltaoneP-2\epsoneP^{2} 
+(2D-3) \epsoneP\epstwoP\nonumber \\ & & +3\epsoneP\deltaoneP +
\epstwoP\deltaoneP +D \epstwoP\epsthreeP-\deltaoneP^{2} 
\nonumber \\ & & + 2D
\deltaoneP\deltatwoP \, .
\end{eqnarray}
The claim made before can now be explicitly checked. The above formula
coincides at first order with Eq.~(\ref{eq:ns}) but differs at second
order. It is interesting to notice that the differences show up only in
those terms containing the constant $D$.

\par

A slightly different set of parameters
$\left(\epsilon,s,\eta,\rho,^2\lambda\right)$ was recently introduced in
Ref.~\cite{Kinney:2007ag}, and expressed in terms of our
$\left(\epsilon_{i},\delta_{i}\right)$ these parameters read
\begin{eqnarray}
  \epsilon&=&\epsilon_{1},\label{eq:kinneyeps}\\
  \eta&=&\epsilon_{1}-\frac{1}{2}(\epsilon_{2}+\delta_{1}),\\
  s&=&-\delta_{1},\\
  \rho&=&\frac{\delta_{1}}{2\epsilon_{1}}\left(2\delta_{2}+3\delta_{1}
    - \epsilon_{2}\right),\\
  ^{2}\lambda&=&\epsilon_{1}^{2}+\frac{1}{2}\big(\epsilon_{2}\epsilon_{3}
  -3\epsilon_{1}\epsilon_{2}+\epsilon_{2}\delta_{1}-3\epsilon_{1}\delta_{1}
  \nonumber\\
  &&+\delta_{1}^{2}+\delta_{1}\delta_{2}\big).\label{eq:kinneyrho}
\end{eqnarray}
Using this correspondance, one can explicitly check that Eq.~(84) of
Ref.~\cite{Kinney:2007ag} matches exactly with our Eq.~(\ref{eq:ns}),
the constant $C$ in that reference being related to the constant $D$
used in the present paper by $D=(C-3)/4$. The small difference in the
actual numerical values of $(C-3)/4=\ln 2 + \gamma_\mathrm{Euler}
-7/4\simeq -0.73$ and our $D=1/3 -\ln 3\simeq -0.76$ is the just the
well-known difference between the WKB approximation versus the first
order integration of the Bessel equation~\cite{Martin:2002vn}.

\section{Cosmological perturbations from DBI inflation}
\label{app:pert}

In this Appendix, we briefly establish the conservation law for
cosmological perturbations in k-inflation. This is important because as
in the standard case~\cite{Martin:1997zd}, this allows us to propagate
the primordial power spectra to the recombination epoch. We also briefly
recall how the equation of motion for the Mukhanov-Sasaki variable can
be obtained.

\par

In the longitudinal gauge, the perturbed FLRW metric reads
\begin{align}
\label{metricgi}
\dd s^2 & =  a^2(\eta )\bigl\{-\left(1+2\Phi \right) \dd \eta
^2 \nonumber \\ & + \left[\left(1-2\Psi
  \right) \delta _{ij} +  h_{ij} \right] \dd
x^i \dd x^j \bigr\} ,
\end{align}
where $\Phi$ and $\Psi$ are the Bardeen potentials which are coupled
through the perturbed Einstein equation to the field perturbations
$\delta\varphi$. The transverse and traceless tensor $h_{ij}$, \ie
satisfying $h_i{}^i=\partial ^jh_{ij}=0$ represents the spin two
fluctuations which are not considered in the following since
they are not affected by the DBI dynamics.

\par

At the first order, the perturbed energy momentum tensor is obtained
from Eq.~(\ref{eq:action}) and its components read
\begin{align}
\delta T^0{}_0 &= -\frac{\gamma ^3}{a^2}\varphi '\delta \varphi '
-\frac{{\rm d}V}{{\rm d}\varphi}\delta \varphi
+\frac{(2+\gamma )(\gamma -1)^2}{2}\frac{{\rm d}T}{{\rm d}\varphi}
\delta \varphi\nonumber \\  & +T\Phi \gamma \left(\gamma ^2-1\right) ,
\\
\delta T^k{}_{0}& = \frac{\gamma }{a^2}\varphi '\delta ^{k\ell}
\partial _{\ell }\delta \varphi ,
\\
\delta T^k{}_{\ell } & = \Biggl[\frac{\gamma }{a^2}\varphi '\delta \varphi '
-\frac{{\rm d}V}{{\rm d}\varphi}\delta \varphi
-\frac{(\gamma -1)^2}{2\gamma}\frac{{\rm d}T}{{\rm d}\varphi}
\delta \varphi \nonumber \\ 
& -T\Phi \frac{\gamma ^2-1}{\gamma}
\Biggr]\delta ^k{}_{\ell}\, .
\end{align}
The gravitational sector remaining standard General Relativity whose
perturbed Einstein tensor can be found in
Ref.~\cite{Mukhanov:1990me}. In particular, one still has the relation
$\Psi=\Phi$ and there is only one degree of freedom because $\Phi$ and
$\delta \varphi $ are related by the perturbed Einstein equations. One
can therefore reduce the study of the scalar sector to the study of a
single variable, as the comoving curvature perturbation
\begin{equation}
\label{eq:zetadef}
\zeta \equiv \Phi + \dfrac{\calH}{\calH^2-\calH'} 
\left(\Phi' + \calH \Phi \right).
\end{equation}
{}From the background and perturbed equations, one can show that the
comoving curvature perturbation can be simplified to
\begin{equation}
\zeta = \Phi  + \calH \dfrac{\delta \varphi}{\varphi '},
\end{equation}
which also matches with its usual expression in standard
inflation. Straightforward manipulations allows us to derive the
expression of the derivative of $\zeta$ and one gets
\begin{equation}
\zeta '=-\frac{\calH}{\calH^2-\calH'}\frac{k^2}{\gamma ^2}\Phi.
\end{equation}
As a result, $\zeta $ is a conserved quantity on scales larger than the
sonic horizon and allows us to propagate the spectrum from horizon exit
till the beginning of the radiation dominated era. As usual, this result
applies if the decaying mode is neglected and in absence of entropy
(isocurvature) perturbations~\cite{Martin:1997zd}.

\par

The Mukhanov-Sasaki variable is now related to the comoving curvature
perturbations by $v_{\mathbf{k}}=z \zeta$ where
\begin{equation}
z= \gamma^{3/2} \dfrac{a \varphi'}{\calH}\,,
\end{equation}
has a $\gamma$ dependence. Using the previous equations, $v_{\bf k}$
is found to obeys the mode equation
\begin{equation}
v_{\bf k}''+\left(\frac{k^{2}}{\gamma^{2}}-\frac{z''}{z}\right)v_{\bf k}=0 .
\end{equation}
This is the equation already derived in
Refs.~\cite{Bean:2007hc,Kinney:2007ag} and used in the text with the
correspondence $\cs =1/\gamma$.

\end{appendix}

\bibliography{references}

\begin{thebibliography}{59}
\expandafter\ifx\csname natexlab\endcsname\relax\def\natexlab#1{#1}\fi
\expandafter\ifx\csname bibnamefont\endcsname\relax
  \def\bibnamefont#1{#1}\fi
\expandafter\ifx\csname bibfnamefont\endcsname\relax
  \def\bibfnamefont#1{#1}\fi
\expandafter\ifx\csname citenamefont\endcsname\relax
  \def\citenamefont#1{#1}\fi
\expandafter\ifx\csname url\endcsname\relax
  \def\url#1{\texttt{#1}}\fi
\expandafter\ifx\csname urlprefix\endcsname\relax\def\urlprefix{URL }\fi
\providecommand{\bibinfo}[2]{#2}
\providecommand{\eprint}[2][]{\url{#2}}

\bibitem[{\citenamefont{Gold et~al.}(2008)}]{Gold:2008kp}
\bibinfo{author}{\bibfnamefont{B.}~\bibnamefont{Gold}} \bibnamefont{et~al.}
  (\bibinfo{collaboration}{WMAP}) (\bibinfo{year}{2008}), \eprint{0803.0715}.

\bibitem[{\citenamefont{Hill et~al.}(2008)}]{Hill:2008hx}
\bibinfo{author}{\bibfnamefont{R.~S.} \bibnamefont{Hill}} \bibnamefont{et~al.}
  (\bibinfo{collaboration}{WMAP}) (\bibinfo{year}{2008}), \eprint{0803.0570}.

\bibitem[{\citenamefont{Hinshaw et~al.}(2008)}]{Hinshaw:2008kr}
\bibinfo{author}{\bibfnamefont{G.}~\bibnamefont{Hinshaw}} \bibnamefont{et~al.}
  (\bibinfo{collaboration}{WMAP}) (\bibinfo{year}{2008}), \eprint{0803.0732}.

\bibitem[{\citenamefont{Nolta et~al.}(2008)}]{Nolta:2008ih}
\bibinfo{author}{\bibfnamefont{M.~R.} \bibnamefont{Nolta}} \bibnamefont{et~al.}
  (\bibinfo{collaboration}{WMAP}) (\bibinfo{year}{2008}), \eprint{0803.0593}.

\bibitem[{\citenamefont{Dunkley et~al.}(2008)}]{Dunkley:2008ie}
\bibinfo{author}{\bibfnamefont{J.}~\bibnamefont{Dunkley}} \bibnamefont{et~al.}
  (\bibinfo{collaboration}{WMAP}) (\bibinfo{year}{2008}), \eprint{0803.0586}.

\bibitem[{\citenamefont{Komatsu et~al.}(2008)}]{Komatsu:2008hk}
\bibinfo{author}{\bibfnamefont{E.}~\bibnamefont{Komatsu}} \bibnamefont{et~al.}
  (\bibinfo{collaboration}{WMAP}) (\bibinfo{year}{2008}), \eprint{0803.0547}.

\bibitem[{\citenamefont{Stewart and Lyth}(1993)}]{Stewart:1993bc}
\bibinfo{author}{\bibfnamefont{E.~D.} \bibnamefont{Stewart}} \bibnamefont{and}
  \bibinfo{author}{\bibfnamefont{D.~H.} \bibnamefont{Lyth}},
  \bibinfo{journal}{Phys. Lett.} \textbf{\bibinfo{volume}{B302}},
  \bibinfo{pages}{171} (\bibinfo{year}{1993}), \eprint{gr-qc/9302019}.

\bibitem[{\citenamefont{Martin and Schwarz}(2000)}]{Martin:1999wa}
\bibinfo{author}{\bibfnamefont{J.}~\bibnamefont{Martin}} \bibnamefont{and}
  \bibinfo{author}{\bibfnamefont{D.~J.} \bibnamefont{Schwarz}},
  \bibinfo{journal}{Phys. Rev.} \textbf{\bibinfo{volume}{D62}},
  \bibinfo{pages}{103520} (\bibinfo{year}{2000}), \eprint{astro-ph/9911225}.

\bibitem[{\citenamefont{Martin et~al.}(2000)\citenamefont{Martin, Riazuelo, and
  Schwarz}}]{Martin:2000ak}
\bibinfo{author}{\bibfnamefont{J.}~\bibnamefont{Martin}},
  \bibinfo{author}{\bibfnamefont{A.}~\bibnamefont{Riazuelo}}, \bibnamefont{and}
  \bibinfo{author}{\bibfnamefont{D.~J.} \bibnamefont{Schwarz}},
  \bibinfo{journal}{Astrophys. J.} \textbf{\bibinfo{volume}{543}},
  \bibinfo{pages}{L99} (\bibinfo{year}{2000}), \eprint{astro-ph/0006392}.

\bibitem[{\citenamefont{Schwarz et~al.}(2001)\citenamefont{Schwarz,
  Terrero-Escalante, and Garcia}}]{Schwarz:2001vv}
\bibinfo{author}{\bibfnamefont{D.~J.} \bibnamefont{Schwarz}},
  \bibinfo{author}{\bibfnamefont{C.~A.} \bibnamefont{Terrero-Escalante}},
  \bibnamefont{and} \bibinfo{author}{\bibfnamefont{A.~A.}
  \bibnamefont{Garcia}}, \bibinfo{journal}{Phys. Lett.}
  \textbf{\bibinfo{volume}{B517}}, \bibinfo{pages}{243} (\bibinfo{year}{2001}),
  \eprint{astro-ph/0106020}.

\bibitem[{\citenamefont{Leach et~al.}(2002)\citenamefont{Leach, Liddle, Martin,
  and Schwarz}}]{Leach:2002ar}
\bibinfo{author}{\bibfnamefont{S.~M.} \bibnamefont{Leach}},
  \bibinfo{author}{\bibfnamefont{A.~R.} \bibnamefont{Liddle}},
  \bibinfo{author}{\bibfnamefont{J.}~\bibnamefont{Martin}}, \bibnamefont{and}
  \bibinfo{author}{\bibfnamefont{D.~J.} \bibnamefont{Schwarz}},
  \bibinfo{journal}{Phys. Rev.} \textbf{\bibinfo{volume}{D66}},
  \bibinfo{pages}{023515} (\bibinfo{year}{2002}), \eprint{astro-ph/0202094}.

\bibitem[{\citenamefont{Schwarz and Terrero-Escalante}(2004)}]{Schwarz:2004tz}
\bibinfo{author}{\bibfnamefont{D.~J.} \bibnamefont{Schwarz}} \bibnamefont{and}
  \bibinfo{author}{\bibfnamefont{C.~A.} \bibnamefont{Terrero-Escalante}},
  \bibinfo{journal}{JCAP} \textbf{\bibinfo{volume}{0408}}, \bibinfo{pages}{003}
  (\bibinfo{year}{2004}), \eprint{hep-ph/0403129}.

\bibitem[{\citenamefont{Lyth and Riotto}(1999)}]{Lyth:1998xn}
\bibinfo{author}{\bibfnamefont{D.~H.} \bibnamefont{Lyth}} \bibnamefont{and}
  \bibinfo{author}{\bibfnamefont{A.}~\bibnamefont{Riotto}},
  \bibinfo{journal}{Phys. Rept.} \textbf{\bibinfo{volume}{314}},
  \bibinfo{pages}{1} (\bibinfo{year}{1999}), \eprint{hep-ph/9807278}.

\bibitem[{\citenamefont{Cline}(2006)}]{Cline:2006hu}
\bibinfo{author}{\bibfnamefont{J.~M.} \bibnamefont{Cline}}
  (\bibinfo{year}{2006}), \eprint{hep-th/0612129}.

\bibitem[{\citenamefont{Kallosh}(2008)}]{Kallosh:2007ig}
\bibinfo{author}{\bibfnamefont{R.}~\bibnamefont{Kallosh}},
  \bibinfo{journal}{Lect. Notes Phys.} \textbf{\bibinfo{volume}{738}},
  \bibinfo{pages}{119} (\bibinfo{year}{2008}), \eprint{hep-th/0702059}.

\bibitem[{\citenamefont{McAllister and Silverstein}(2008)}]{McAllister:2007bg}
\bibinfo{author}{\bibfnamefont{L.}~\bibnamefont{McAllister}} \bibnamefont{and}
  \bibinfo{author}{\bibfnamefont{E.}~\bibnamefont{Silverstein}},
  \bibinfo{journal}{Gen. Rel. Grav.} \textbf{\bibinfo{volume}{40}},
  \bibinfo{pages}{565} (\bibinfo{year}{2008}), \eprint{0710.2951}.

\bibitem[{\citenamefont{Armendariz-Picon
  et~al.}(1999)\citenamefont{Armendariz-Picon, Damour, and
  Mukhanov}}]{ArmendarizPicon:1999rj}
\bibinfo{author}{\bibfnamefont{C.}~\bibnamefont{Armendariz-Picon}},
  \bibinfo{author}{\bibfnamefont{T.}~\bibnamefont{Damour}}, \bibnamefont{and}
  \bibinfo{author}{\bibfnamefont{V.~F.} \bibnamefont{Mukhanov}},
  \bibinfo{journal}{Phys. Lett.} \textbf{\bibinfo{volume}{B458}},
  \bibinfo{pages}{209} (\bibinfo{year}{1999}), \eprint{hep-th/9904075}.

\bibitem[{\citenamefont{Garriga and Mukhanov}(1999)}]{Garriga:1999vw}
\bibinfo{author}{\bibfnamefont{J.}~\bibnamefont{Garriga}} \bibnamefont{and}
  \bibinfo{author}{\bibfnamefont{V.~F.} \bibnamefont{Mukhanov}},
  \bibinfo{journal}{Phys. Lett.} \textbf{\bibinfo{volume}{B458}},
  \bibinfo{pages}{219} (\bibinfo{year}{1999}), \eprint{hep-th/9904176}.

\bibitem[{\citenamefont{Kachru et~al.}(2003)}]{Kachru:2003sx}
\bibinfo{author}{\bibfnamefont{S.}~\bibnamefont{Kachru}} \bibnamefont{et~al.},
  \bibinfo{journal}{JCAP} \textbf{\bibinfo{volume}{0310}}, \bibinfo{pages}{013}
  (\bibinfo{year}{2003}), \eprint{hep-th/0308055}.

\bibitem[{\citenamefont{Alishahiha et~al.}(2004)\citenamefont{Alishahiha,
  Silverstein, and Tong}}]{Alishahiha:2004eh}
\bibinfo{author}{\bibfnamefont{M.}~\bibnamefont{Alishahiha}},
  \bibinfo{author}{\bibfnamefont{E.}~\bibnamefont{Silverstein}},
  \bibnamefont{and} \bibinfo{author}{\bibfnamefont{D.}~\bibnamefont{Tong}},
  \bibinfo{journal}{Phys. Rev.} \textbf{\bibinfo{volume}{D70}},
  \bibinfo{pages}{123505} (\bibinfo{year}{2004}), \eprint{hep-th/0404084}.

\bibitem[{\citenamefont{Dvali and Tye}(1999)}]{Dvali:1998pa}
\bibinfo{author}{\bibfnamefont{G.~R.} \bibnamefont{Dvali}} \bibnamefont{and}
  \bibinfo{author}{\bibfnamefont{S.~H.~H.} \bibnamefont{Tye}},
  \bibinfo{journal}{Phys. Lett.} \textbf{\bibinfo{volume}{B450}},
  \bibinfo{pages}{72} (\bibinfo{year}{1999}), \eprint{hep-ph/9812483}.

\bibitem[{\citenamefont{Dvali et~al.}(2001)\citenamefont{Dvali, Shafi, and
  Solganik}}]{Dvali:2001fw}
\bibinfo{author}{\bibfnamefont{G.~R.} \bibnamefont{Dvali}},
  \bibinfo{author}{\bibfnamefont{Q.}~\bibnamefont{Shafi}}, \bibnamefont{and}
  \bibinfo{author}{\bibfnamefont{S.}~\bibnamefont{Solganik}}
  (\bibinfo{year}{2001}), \eprint{hep-th/0105203}.

\bibitem[{\citenamefont{Alexander}(2001)}]{Alexander:2001ks}
\bibinfo{author}{\bibfnamefont{S.~H.~S.} \bibnamefont{Alexander}},
  \bibinfo{journal}{Phys. Rev.} \textbf{\bibinfo{volume}{D65}},
  \bibinfo{pages}{023507} (\bibinfo{year}{2001}), \eprint{hep-th/0105032}.

\bibitem[{\citenamefont{Baumann et~al.}(2008)\citenamefont{Baumann, Dymarsky,
  Klebanov, and McAllister}}]{Baumann:2007ah}
\bibinfo{author}{\bibfnamefont{D.}~\bibnamefont{Baumann}},
  \bibinfo{author}{\bibfnamefont{A.}~\bibnamefont{Dymarsky}},
  \bibinfo{author}{\bibfnamefont{I.~R.} \bibnamefont{Klebanov}},
  \bibnamefont{and}
  \bibinfo{author}{\bibfnamefont{L.}~\bibnamefont{McAllister}},
  \bibinfo{journal}{JCAP} \textbf{\bibinfo{volume}{0801}}, \bibinfo{pages}{024}
  (\bibinfo{year}{2008}), \eprint{0706.0360}.

\bibitem[{\citenamefont{Baumann et~al.}(2007)\citenamefont{Baumann, Dymarsky,
  Klebanov, McAllister, and Steinhardt}}]{Baumann:2007np}
\bibinfo{author}{\bibfnamefont{D.}~\bibnamefont{Baumann}},
  \bibinfo{author}{\bibfnamefont{A.}~\bibnamefont{Dymarsky}},
  \bibinfo{author}{\bibfnamefont{I.~R.} \bibnamefont{Klebanov}},
  \bibinfo{author}{\bibfnamefont{L.}~\bibnamefont{McAllister}},
  \bibnamefont{and} \bibinfo{author}{\bibfnamefont{P.~J.}
  \bibnamefont{Steinhardt}}, \bibinfo{journal}{Phys. Rev. Lett.}
  \textbf{\bibinfo{volume}{99}}, \bibinfo{pages}{141601}
  (\bibinfo{year}{2007}), \eprint{0705.3837}.

\bibitem[{\citenamefont{Chen et~al.}(2007)\citenamefont{Chen, Huang, Kachru,
  and Shiu}}]{Chen:2006nt}
\bibinfo{author}{\bibfnamefont{X.}~\bibnamefont{Chen}},
  \bibinfo{author}{\bibfnamefont{M.-x.} \bibnamefont{Huang}},
  \bibinfo{author}{\bibfnamefont{S.}~\bibnamefont{Kachru}}, \bibnamefont{and}
  \bibinfo{author}{\bibfnamefont{G.}~\bibnamefont{Shiu}},
  \bibinfo{journal}{JCAP} \textbf{\bibinfo{volume}{0701}}, \bibinfo{pages}{002}
  (\bibinfo{year}{2007}), \eprint{hep-th/0605045}.

\bibitem[{\citenamefont{Kinney and Tzirakis}(2008)}]{Kinney:2007ag}
\bibinfo{author}{\bibfnamefont{W.~H.} \bibnamefont{Kinney}} \bibnamefont{and}
  \bibinfo{author}{\bibfnamefont{K.}~\bibnamefont{Tzirakis}},
  \bibinfo{journal}{Phys. Rev.} \textbf{\bibinfo{volume}{D77}},
  \bibinfo{pages}{103517} (\bibinfo{year}{2008}), \eprint{0712.2043}.

\bibitem[{\citenamefont{Peiris et~al.}(2007)\citenamefont{Peiris, Baumann,
  Friedman, and Cooray}}]{Peiris:2007gz}
\bibinfo{author}{\bibfnamefont{H.~V.} \bibnamefont{Peiris}},
  \bibinfo{author}{\bibfnamefont{D.}~\bibnamefont{Baumann}},
  \bibinfo{author}{\bibfnamefont{B.}~\bibnamefont{Friedman}}, \bibnamefont{and}
  \bibinfo{author}{\bibfnamefont{A.}~\bibnamefont{Cooray}},
  \bibinfo{journal}{Phys. Rev.} \textbf{\bibinfo{volume}{D76}},
  \bibinfo{pages}{103517} (\bibinfo{year}{2007}), \eprint{0706.1240}.

\bibitem[{\citenamefont{Habib et~al.}(2002)\citenamefont{Habib, Heitmann,
  Jungman, and Molina-Paris}}]{Habib:2002yi}
\bibinfo{author}{\bibfnamefont{S.}~\bibnamefont{Habib}},
  \bibinfo{author}{\bibfnamefont{K.}~\bibnamefont{Heitmann}},
  \bibinfo{author}{\bibfnamefont{G.}~\bibnamefont{Jungman}}, \bibnamefont{and}
  \bibinfo{author}{\bibfnamefont{C.}~\bibnamefont{Molina-Paris}},
  \bibinfo{journal}{Phys. Rev. Lett.} \textbf{\bibinfo{volume}{89}},
  \bibinfo{pages}{281301} (\bibinfo{year}{2002}), \eprint{astro-ph/0208443}.

\bibitem[{\citenamefont{Habib et~al.}(2004)\citenamefont{Habib, Heinen,
  Heitmann, Jungman, and Molina-Paris}}]{Habib:2004kc}
\bibinfo{author}{\bibfnamefont{S.}~\bibnamefont{Habib}},
  \bibinfo{author}{\bibfnamefont{A.}~\bibnamefont{Heinen}},
  \bibinfo{author}{\bibfnamefont{K.}~\bibnamefont{Heitmann}},
  \bibinfo{author}{\bibfnamefont{G.}~\bibnamefont{Jungman}}, \bibnamefont{and}
  \bibinfo{author}{\bibfnamefont{C.}~\bibnamefont{Molina-Paris}},
  \bibinfo{journal}{Phys. Rev.} \textbf{\bibinfo{volume}{D70}},
  \bibinfo{pages}{083507} (\bibinfo{year}{2004}), \eprint{astro-ph/0406134}.

\bibitem[{\citenamefont{Spalinski}(2007{\natexlab{a}})}]{Spalinski:2007dv}
\bibinfo{author}{\bibfnamefont{M.}~\bibnamefont{Spalinski}},
  \bibinfo{journal}{JCAP} \textbf{\bibinfo{volume}{0705}}, \bibinfo{pages}{017}
  (\bibinfo{year}{2007}{\natexlab{a}}), \eprint{hep-th/0702196}.

\bibitem[{\citenamefont{Spalinski}(2007{\natexlab{b}})}]{Spalinski:2007qy}
\bibinfo{author}{\bibfnamefont{M.}~\bibnamefont{Spalinski}},
  \bibinfo{journal}{Phys. Lett.} \textbf{\bibinfo{volume}{B650}},
  \bibinfo{pages}{313} (\bibinfo{year}{2007}{\natexlab{b}}),
  \eprint{hep-th/0703248}.

\bibitem[{\citenamefont{Klebanov and Strassler}(2000)}]{Klebanov:2000hb}
\bibinfo{author}{\bibfnamefont{I.~R.} \bibnamefont{Klebanov}} \bibnamefont{and}
  \bibinfo{author}{\bibfnamefont{M.~J.} \bibnamefont{Strassler}},
  \bibinfo{journal}{JHEP} \textbf{\bibinfo{volume}{08}}, \bibinfo{pages}{052}
  (\bibinfo{year}{2000}), \eprint{hep-th/0007191}.

\bibitem[{\citenamefont{Bean et~al.}(2007)\citenamefont{Bean, Shandera,
  Henry~Tye, and Xu}}]{Bean:2007hc}
\bibinfo{author}{\bibfnamefont{R.}~\bibnamefont{Bean}},
  \bibinfo{author}{\bibfnamefont{S.~E.} \bibnamefont{Shandera}},
  \bibinfo{author}{\bibfnamefont{S.~H.} \bibnamefont{Henry~Tye}},
  \bibnamefont{and} \bibinfo{author}{\bibfnamefont{J.}~\bibnamefont{Xu}},
  \bibinfo{journal}{JCAP} \textbf{\bibinfo{volume}{0705}}, \bibinfo{pages}{004}
  (\bibinfo{year}{2007}), \eprint{hep-th/0702107}.

\bibitem[{\citenamefont{Shandera and Tye}(2006)}]{Shandera:2006ax}
\bibinfo{author}{\bibfnamefont{S.~E.} \bibnamefont{Shandera}} \bibnamefont{and}
  \bibinfo{author}{\bibfnamefont{S.~H.~H.} \bibnamefont{Tye}},
  \bibinfo{journal}{JCAP} \textbf{\bibinfo{volume}{0605}}, \bibinfo{pages}{007}
  (\bibinfo{year}{2006}), \eprint{hep-th/0601099}.

\bibitem[{\citenamefont{Henry~Tye}(2008)}]{HenryTye:2006uv}
\bibinfo{author}{\bibfnamefont{S.~H.} \bibnamefont{Henry~Tye}},
  \bibinfo{journal}{Lect. Notes Phys.} \textbf{\bibinfo{volume}{737}},
  \bibinfo{pages}{949} (\bibinfo{year}{2008}), \eprint{hep-th/0610221}.

\bibitem[{\citenamefont{Burgess}(2006)}]{Burgess:2007pz}
\bibinfo{author}{\bibfnamefont{C.~P.} \bibnamefont{Burgess}},
  \bibinfo{journal}{PoS} \textbf{\bibinfo{volume}{P2GC}}, \bibinfo{pages}{008}
  (\bibinfo{year}{2006}), \eprint{0708.2865}.

\bibitem[{\citenamefont{Bruneton and Esposito-Farese}(2007)}]{Bruneton:2007si}
\bibinfo{author}{\bibfnamefont{J.-P.} \bibnamefont{Bruneton}} \bibnamefont{and}
  \bibinfo{author}{\bibfnamefont{G.}~\bibnamefont{Esposito-Farese}},
  \bibinfo{journal}{Phys. Rev.} \textbf{\bibinfo{volume}{D76}},
  \bibinfo{pages}{124012} (\bibinfo{year}{2007}), \eprint{0705.4043}.

\bibitem[{\citenamefont{Silverstein and Tong}(2004)}]{Silverstein:2003hf}
\bibinfo{author}{\bibfnamefont{E.}~\bibnamefont{Silverstein}} \bibnamefont{and}
  \bibinfo{author}{\bibfnamefont{D.}~\bibnamefont{Tong}},
  \bibinfo{journal}{Phys. Rev.} \textbf{\bibinfo{volume}{D70}},
  \bibinfo{pages}{103505} (\bibinfo{year}{2004}), \eprint{hep-th/0310221}.

\bibitem[{\citenamefont{Lorenz et~al.}(2008{\natexlab{a}})\citenamefont{Lorenz,
  Martin, and Ringeval}}]{Lorenz:2007ze}
\bibinfo{author}{\bibfnamefont{L.}~\bibnamefont{Lorenz}},
  \bibinfo{author}{\bibfnamefont{J.}~\bibnamefont{Martin}}, \bibnamefont{and}
  \bibinfo{author}{\bibfnamefont{C.}~\bibnamefont{Ringeval}},
  \bibinfo{journal}{JCAP} \textbf{\bibinfo{volume}{0804}}, \bibinfo{pages}{001}
  (\bibinfo{year}{2008}{\natexlab{a}}), \eprint{0709.3758}.

\bibitem[{\citenamefont{Bean et~al.}(2008{\natexlab{a}})\citenamefont{Bean,
  Chung, and Geshnizjani}}]{Bean:2008ga}
\bibinfo{author}{\bibfnamefont{R.}~\bibnamefont{Bean}},
  \bibinfo{author}{\bibfnamefont{D.~J.~H.} \bibnamefont{Chung}},
  \bibnamefont{and}
  \bibinfo{author}{\bibfnamefont{G.}~\bibnamefont{Geshnizjani}}
  (\bibinfo{year}{2008}{\natexlab{a}}), \eprint{0801.0742}.

\bibitem[{\citenamefont{Bean et~al.}(2008{\natexlab{b}})\citenamefont{Bean,
  Chen, Peiris, and Xu}}]{Bean:2007eh}
\bibinfo{author}{\bibfnamefont{R.}~\bibnamefont{Bean}},
  \bibinfo{author}{\bibfnamefont{X.}~\bibnamefont{Chen}},
  \bibinfo{author}{\bibfnamefont{H.~V.} \bibnamefont{Peiris}},
  \bibnamefont{and} \bibinfo{author}{\bibfnamefont{J.}~\bibnamefont{Xu}},
  \bibinfo{journal}{Phys. Rev.} \textbf{\bibinfo{volume}{D77}},
  \bibinfo{pages}{023527} (\bibinfo{year}{2008}{\natexlab{b}}),
  \eprint{0710.1812}.

\bibitem[{\citenamefont{Martin and Schwarz}(2003)}]{Martin:2002vn}
\bibinfo{author}{\bibfnamefont{J.}~\bibnamefont{Martin}} \bibnamefont{and}
  \bibinfo{author}{\bibfnamefont{D.~J.} \bibnamefont{Schwarz}},
  \bibinfo{journal}{Phys. Rev.} \textbf{\bibinfo{volume}{D67}},
  \bibinfo{pages}{083512} (\bibinfo{year}{2003}), \eprint{astro-ph/0210090}.

\bibitem[{\citenamefont{Casadio et~al.}(2005)\citenamefont{Casadio, Finelli,
  Luzzi, and Venturi}}]{Casadio:2004ru}
\bibinfo{author}{\bibfnamefont{R.}~\bibnamefont{Casadio}},
  \bibinfo{author}{\bibfnamefont{F.}~\bibnamefont{Finelli}},
  \bibinfo{author}{\bibfnamefont{M.}~\bibnamefont{Luzzi}}, \bibnamefont{and}
  \bibinfo{author}{\bibfnamefont{G.}~\bibnamefont{Venturi}},
  \bibinfo{journal}{Phys. Rev.} \textbf{\bibinfo{volume}{D71}},
  \bibinfo{pages}{043517} (\bibinfo{year}{2005}), \eprint{gr-qc/0410092}.

\bibitem[{\citenamefont{Lorenz et~al.}(2008{\natexlab{b}})\citenamefont{Lorenz,
  Martin, and Ringeval}}]{Lorenz:2008je}
\bibinfo{author}{\bibfnamefont{L.}~\bibnamefont{Lorenz}},
  \bibinfo{author}{\bibfnamefont{J.}~\bibnamefont{Martin}}, \bibnamefont{and}
  \bibinfo{author}{\bibfnamefont{C.}~\bibnamefont{Ringeval}}
  (\bibinfo{year}{2008}{\natexlab{b}}), \eprint{0807.2414}.

\bibitem[{\citenamefont{Spalinski}(2008)}]{Spalinski:2007un}
\bibinfo{author}{\bibfnamefont{M.}~\bibnamefont{Spalinski}},
  \bibinfo{journal}{JCAP} \textbf{\bibinfo{volume}{0804}}, \bibinfo{pages}{002}
  (\bibinfo{year}{2008}), \eprint{0711.4326}.

\bibitem[{\citenamefont{Chen}(2005{\natexlab{a}})}]{Chen:2004gc}
\bibinfo{author}{\bibfnamefont{X.}~\bibnamefont{Chen}}, \bibinfo{journal}{Phys.
  Rev.} \textbf{\bibinfo{volume}{D71}}, \bibinfo{pages}{063506}
  (\bibinfo{year}{2005}{\natexlab{a}}), \eprint{hep-th/0408084}.

\bibitem[{\citenamefont{Chen}(2005{\natexlab{b}})}]{Chen:2005ad}
\bibinfo{author}{\bibfnamefont{X.}~\bibnamefont{Chen}}, \bibinfo{journal}{JHEP}
  \textbf{\bibinfo{volume}{08}}, \bibinfo{pages}{045}
  (\bibinfo{year}{2005}{\natexlab{b}}), \eprint{hep-th/0501184}.

\bibitem[{\citenamefont{Bean et~al.}(2006)\citenamefont{Bean, Dunkley, and
  Pierpaoli}}]{Bean:2006qz}
\bibinfo{author}{\bibfnamefont{R.}~\bibnamefont{Bean}},
  \bibinfo{author}{\bibfnamefont{J.}~\bibnamefont{Dunkley}}, \bibnamefont{and}
  \bibinfo{author}{\bibfnamefont{E.}~\bibnamefont{Pierpaoli}},
  \bibinfo{journal}{Phys. Rev.} \textbf{\bibinfo{volume}{D74}},
  \bibinfo{pages}{063503} (\bibinfo{year}{2006}), \eprint{astro-ph/0606685}.

\bibitem[{\citenamefont{Chen et~al.}(2006)\citenamefont{Chen, Sarangi,
  Henry~Tye, and Xu}}]{Chen:2006hs}
\bibinfo{author}{\bibfnamefont{X.}~\bibnamefont{Chen}},
  \bibinfo{author}{\bibfnamefont{S.}~\bibnamefont{Sarangi}},
  \bibinfo{author}{\bibfnamefont{S.~H.} \bibnamefont{Henry~Tye}},
  \bibnamefont{and} \bibinfo{author}{\bibfnamefont{J.}~\bibnamefont{Xu}},
  \bibinfo{journal}{JCAP} \textbf{\bibinfo{volume}{0611}}, \bibinfo{pages}{015}
  (\bibinfo{year}{2006}), \eprint{hep-th/0608082}.

\bibitem[{\citenamefont{Easson et~al.}(2008)\citenamefont{Easson, Gregory,
  Mota, Tasinato, and Zavala}}]{Easson:2007dh}
\bibinfo{author}{\bibfnamefont{D.~A.} \bibnamefont{Easson}},
  \bibinfo{author}{\bibfnamefont{R.}~\bibnamefont{Gregory}},
  \bibinfo{author}{\bibfnamefont{D.~F.} \bibnamefont{Mota}},
  \bibinfo{author}{\bibfnamefont{G.}~\bibnamefont{Tasinato}}, \bibnamefont{and}
  \bibinfo{author}{\bibfnamefont{I.}~\bibnamefont{Zavala}},
  \bibinfo{journal}{JCAP} \textbf{\bibinfo{volume}{0802}}, \bibinfo{pages}{010}
  (\bibinfo{year}{2008}), \eprint{0709.2666}.

\bibitem[{\citenamefont{Easson et~al.}(2007)\citenamefont{Easson, Gregory,
  Tasinato, and Zavala}}]{Easson:2007fz}
\bibinfo{author}{\bibfnamefont{D.}~\bibnamefont{Easson}},
  \bibinfo{author}{\bibfnamefont{R.}~\bibnamefont{Gregory}},
  \bibinfo{author}{\bibfnamefont{G.}~\bibnamefont{Tasinato}}, \bibnamefont{and}
  \bibinfo{author}{\bibfnamefont{I.}~\bibnamefont{Zavala}},
  \bibinfo{journal}{JHEP} \textbf{\bibinfo{volume}{04}}, \bibinfo{pages}{026}
  (\bibinfo{year}{2007}), \eprint{hep-th/0701252}.

\bibitem[{\citenamefont{Baumann and McAllister}(2007)}]{Baumann:2006cd}
\bibinfo{author}{\bibfnamefont{D.}~\bibnamefont{Baumann}} \bibnamefont{and}
  \bibinfo{author}{\bibfnamefont{L.}~\bibnamefont{McAllister}},
  \bibinfo{journal}{Phys. Rev.} \textbf{\bibinfo{volume}{D75}},
  \bibinfo{pages}{123508} (\bibinfo{year}{2007}), \eprint{hep-th/0610285}.

\bibitem[{\citenamefont{Martin and Ringeval}(2006)}]{Martin:2006rs}
\bibinfo{author}{\bibfnamefont{J.}~\bibnamefont{Martin}} \bibnamefont{and}
  \bibinfo{author}{\bibfnamefont{C.}~\bibnamefont{Ringeval}},
  \bibinfo{journal}{JCAP} \textbf{\bibinfo{volume}{0608}}, \bibinfo{pages}{009}
  (\bibinfo{year}{2006}), \eprint{astro-ph/0605367}.

\bibitem[{\citenamefont{Kinney et~al.}(2006)\citenamefont{Kinney, Kolb,
  Melchiorri, and Riotto}}]{Kinney:2006qm}
\bibinfo{author}{\bibfnamefont{W.~H.} \bibnamefont{Kinney}},
  \bibinfo{author}{\bibfnamefont{E.~W.} \bibnamefont{Kolb}},
  \bibinfo{author}{\bibfnamefont{A.}~\bibnamefont{Melchiorri}},
  \bibnamefont{and} \bibinfo{author}{\bibfnamefont{A.}~\bibnamefont{Riotto}},
  \bibinfo{journal}{Phys. Rev.} \textbf{\bibinfo{volume}{D74}},
  \bibinfo{pages}{023502} (\bibinfo{year}{2006}), \eprint{astro-ph/0605338}.

\bibitem[{\citenamefont{Seery and Lidsey}(2005)}]{Seery:2005wm}
\bibinfo{author}{\bibfnamefont{D.}~\bibnamefont{Seery}} \bibnamefont{and}
  \bibinfo{author}{\bibfnamefont{J.~E.} \bibnamefont{Lidsey}},
  \bibinfo{journal}{JCAP} \textbf{\bibinfo{volume}{0506}}, \bibinfo{pages}{003}
  (\bibinfo{year}{2005}), \eprint{astro-ph/0503692}.

\bibitem[{\citenamefont{Abramowitz and Stegun}(1970)}]{Abramovitz:1970aa}
\bibinfo{author}{\bibfnamefont{M.}~\bibnamefont{Abramowitz}} \bibnamefont{and}
  \bibinfo{author}{\bibfnamefont{I.~A.} \bibnamefont{Stegun}},
  \emph{\bibinfo{title}{Handbook of mathematical functions with formulas,
  graphs, and mathematical tables}} (\bibinfo{publisher}{National Bureau of
  Standards}, \bibinfo{address}{Washington, US}, \bibinfo{year}{1970}),
  \bibinfo{edition}{ninth} ed.

\bibitem[{\citenamefont{Martin and Schwarz}(1998)}]{Martin:1997zd}
\bibinfo{author}{\bibfnamefont{J.}~\bibnamefont{Martin}} \bibnamefont{and}
  \bibinfo{author}{\bibfnamefont{D.~J.} \bibnamefont{Schwarz}},
  \bibinfo{journal}{Phys. Rev.} \textbf{\bibinfo{volume}{D57}},
  \bibinfo{pages}{3302} (\bibinfo{year}{1998}), \eprint{gr-qc/9704049}.

\bibitem[{\citenamefont{Mukhanov et~al.}(1992)\citenamefont{Mukhanov, Feldman,
  and Brandenberger}}]{Mukhanov:1990me}
\bibinfo{author}{\bibfnamefont{V.~F.} \bibnamefont{Mukhanov}},
  \bibinfo{author}{\bibfnamefont{H.~A.} \bibnamefont{Feldman}},
  \bibnamefont{and} \bibinfo{author}{\bibfnamefont{R.~H.}
  \bibnamefont{Brandenberger}}, \bibinfo{journal}{Phys. Rept.}
  \textbf{\bibinfo{volume}{215}}, \bibinfo{pages}{203} (\bibinfo{year}{1992}).

\end{thebibliography}
\end{document}